\documentclass[12pt]{iopart}
\usepackage{iopams}
\usepackage{graphicx}

\expandafter\let\csname equation*\endcsname\relax
\expandafter\let\csname endequation*\endcsname\relax
\usepackage{amsmath}  
\usepackage{amssymb} 
\usepackage{xcolor} 

\usepackage{hyperref}
\hypersetup{
    colorlinks=true,
    citecolor=blue,
    linkcolor=blue,
    filecolor=blue,      
    urlcolor=blue,
}

\newcommand{\tn}[1]{\textnormal{#1}}
\newcommand{\tnl}[1]{\textnormal{\tiny #1}}

\begin{document}

\title[Direct Detection of Dark Matter]{Direct Detection of WIMP Dark Matter: \\ Concepts and Status}

\author{Marc Schumann}

\address{Physikalisches Institut, Universit\"at Freiburg, 79104 Freiburg, Germany}
\ead{marc.schumann@physik.uni-freiburg.de}
\vspace{10pt}

\begin{abstract}
The existence of dark matter as evidenced by numerous indirect observations is one of the most important indications that there must be physics beyond the Standard Model of particle physics. This article reviews the concepts of direct detection of dark matter in the form of Weakly Interacting Massive Particles (WIMPs) in ultra-sensitive detectors located in underground laboratories, discusses the expected signatures, detector concepts, and how the stringent low-background requirements are achieved. Finally, it summarizes the current status of the field and provides an outlook on the years to come.

\end{abstract}

%
%

%

\section{Introduction: Dark Matter and WIMPs}

More than 95\% of the matter and energy content of the Universe is dark~\cite{Ade:2015xua}, i.e., it does not (or only very weakly) interact with photons and ordinary baryonic matter. A total of 25.9\% is made of dark matter, a not-yet-identified form of matter which builds large scale structures in the Universe. (The remaining 69.1\% are due to dark energy, responsible for the accelerated expansion of the Universe.) Since dark matter is five times more abundant than baryonic matter and thus dominated the evolution of the Universe from the end of the radiation era (at about 70'000\,years after the big bang) until redshifts of $z\gtrsim0.5$ (corresponding to $9.4\times 10^9$\,y), it essentially shaped the Universe that can be observed today. 

However, as of today there is only indirect evidence for the existence of dark matter, based on the observation of gravitational effects. A plethora of experiments aim at the direct detection of dark matter by searching for signals from dark matter particles scattering off Earth-based detectors. This article reviews the concepts of these direct detection experiments, summarizes important aspects for the interpretation of their results and provides an up-to-date status of the field\footnote{Please note that I do not even try to provide a complete list of references in this review; in many cases I decided to refer to recent reviews which provide an up-to-date view on a specific topic, rather than to the original work(s). In other cases I quote particular examples from a generally much longer list of possible references. I also focus mainly on the experiments which are most relevant at the time of writing. This does not imply that a project was not very important at an earlier stage.}. 

In this first section, we will briefly discuss why most scientists are convinced that there is plenty of dark matter in the Universe, how it is distributed in galaxies such as the Milky Way, and will introduce the weakly interacting massive particle (WIMP), one of the most promising candidates for the dark matter particle. The basics of direct detection experiments are presented in Section~\ref{sec::ddbasics}; a discussion of the field's current status follows in Section~\ref{sec::status}.

\subsection{Evidence for Dark Matter}
\label{sec::evidence}

Even though the dark matter particle has not yet been detected directly and it's properties remain largely unknown, there is plenty of indirect evidence from astronomy and cosmology that a large fraction of the matter in the Universe is dark. Since dark matter neither interacts electromagnetically nor strongly all evidence is based on effects of gravity. That it may interact with (sub-)weak-scale cross sections with ordinary matter an assumption, see Section~\ref{sec::wimps}. By now, dark matter was indirectly seen galactic, galaxy-cluster and cosmologic scales.  The amount of dark matter in the solar system is too small to induce observable effects in the motion of planets or spacecrafts~\cite{ref::pitjev}.

The pioneering observations were done by Jan Oort (1932)~\cite{ref::oort} and Fritz Zwicky (1933)~\cite{ref::zwicky}. Oort studied the velocities of stars in the solar neighborhood and found that these are too high to be explained by the luminous mass in the Galaxy. Zwicky applied the virial theorem, which relates the average kinetic and potential energies of a gravitationally bound system, to the Coma galaxy cluster. By estimating the time and mass-averaged galaxy velocity $\langle v^2 \rangle$ in the cluster via redshift measurements, he realized that the cluster mass derived via the virial theorem was much larger than the luminous mass in the cluster. Zwicky attributed the discrepancy to a new form of ``dark matter'' (``Dunkle Materie'' in the article written in German). This early work, however, did not have a significant impact in the scientific community. (A comprehensive review on the history of dark matter can be found in~\cite{Bertone:2016nfn}.)

Since the mid 1960s and starting with Andromeda Galaxy (M31), Vera Rubin studied the rotation curves of spiral galaxies with a sensitive spectrograph built by Kent Ford, focusing on H-II regions of ionized atomic hydrogen at different distances to the galaxy centers. She discovered that the observed rotation velocity $v_r$ did not follow the Keplerian $v_r \propto r^{-\frac{1}{2}}$ decrease expected from the distribution of stars. Instead, all galaxies show a flat (or even slightly increasing) velocity profile after an initial rise attributed to the central bulge~\cite{ref::rubin}. The findings were later confirmed by radio-observations using the 21\,cm line of atomic hydrogen which could show that $v_r$ remains constant even in regions far outside of the visible disk of the galaxy~\cite{ref::bosma}. It is straight forward to show that the flat profile can be explained by a density distribution $\rho(r) \propto r^{-2}$, which is a much more modest decrease than the observed $\rho_s(r) \propto r^{-3.5}$ of the number density in the stellar halo of the Milky Way~\cite{ref::slater}. One can conclude that all galaxies appear to be embedded in a large halo of dark matter.

Further evidence comes from gravitational lensing, where invisible dark matter clumps in the foreground distort the images of luminous objects in the background (for a review, see~\cite{Massey:2010hh}). Among the plethora of lensing systems, a particular iconic example is the Bullet-Cluster~\cite{Clowe:2006eq}: it consists of two galaxy clusters which have traversed themselves. The collision led to a separation of the baryonic matter, which is dominated by hot X-ray emitting plasma and accumulates between the clusters, and the dark matter, which is observed by gravitational lensing. The dissipation-less dark mass component traces the distribution of the cluster's galaxies which also don't undergo collisions. The separation cannot be explained by a modification of the laws of gravity and is thus a clear observational evidence for the existence of dark matter. In its entirety, measurements of gravitational lensing indicate that dark matter is about five times more abundant than ordinary matter and not distributed in massive objects of astronomical size. It moves at non-relativistic velocities (``cold'' dark matter). All interactions other than gravity are at most very small; this also holds for dark matter self-interactions~\cite{Massey:2010hh}.

\begin{figure}[b!]
\centering
\includegraphics[width=0.7\textwidth]{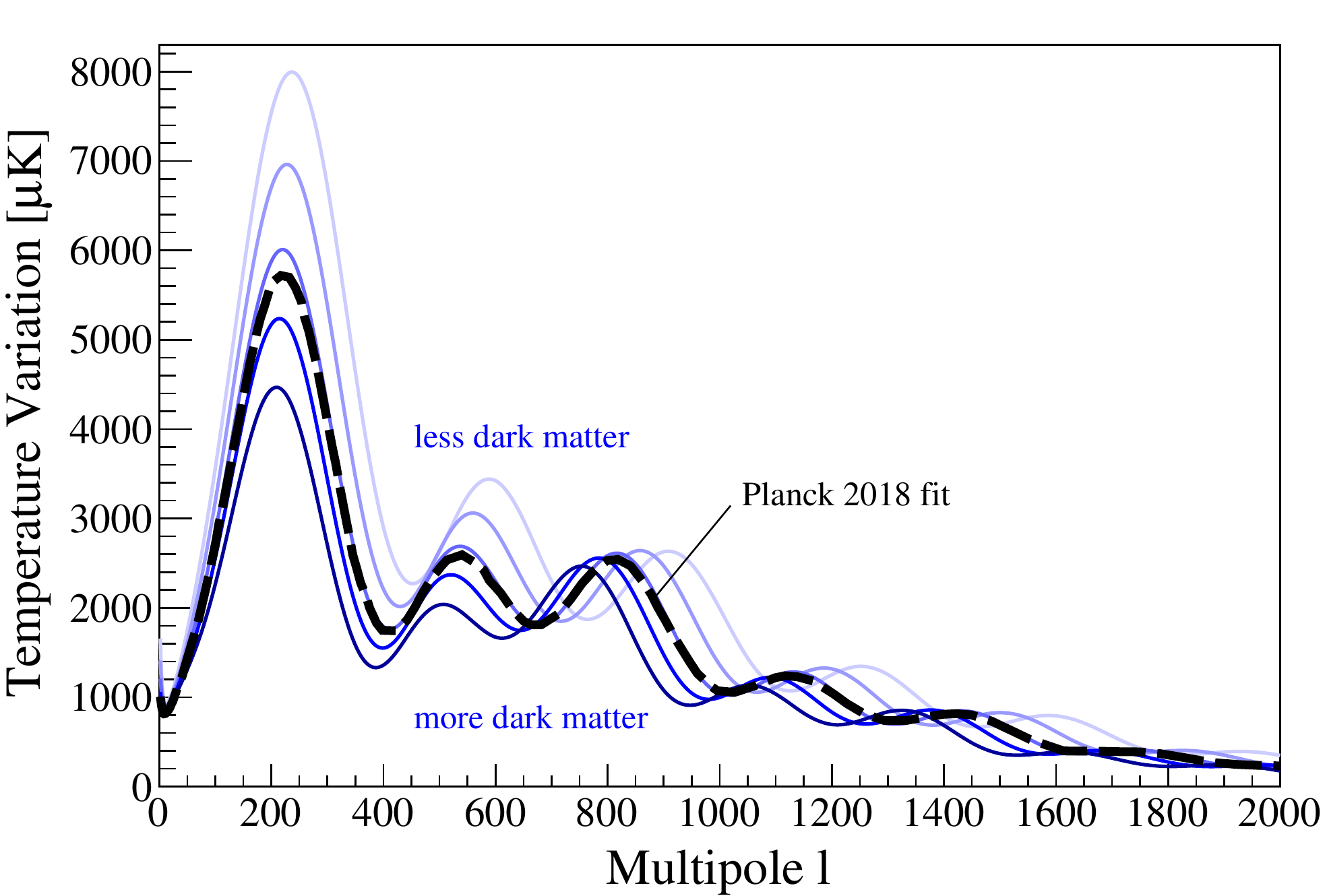}
\caption{Temperature power spectrum of the cosmic microwave background (CMB) for a dark matter density contribution $\Omega_\tnl{CDM}$ varying between 0.11 and and 0.43 (blue lines). All all other input parameters of the model are kept constant. The dashed black lines shows the best fit to the Planck data from the 2018 release~\cite{Aghanim:2018eyx}. Small (large) multipoles $\ell$ correspond to lange (small) angular scales: the main acoustic peak at $\ell \sim 200$ is at $\sim$1$^\circ$; $\ell=1800$ corresponds to $\sim$0.1$^\circ$.} \label{fig::cmb}
\end{figure}

At largest scales, evidence for dark matter comes from the distribution of structure in the Universe~\cite{Percival2001,Eisenstein:2005su} as well as from the precision analysis of the cosmic microwave background (CMB)~\cite{Ade:2015xua,Aghanim:2018eyx}. This almost perfect black body spectrum is a relic of the big bang, created when photons decoupled from the plasma of electrons and protons (at $T\approx 3000$\,K) such that the Universe became optically transparent. The expansion of the Universe has cooled down the radiation to $T_\tnl{CMB}=2.725$\,K. Temperature anisotropies at the $10^{-5}$-level can be observed in the CMB; they originate from ``acoustic'' temperature and density  fluctuations in the early Universe and carry plenty of cosmological information. The power spectrum of the fluctuations shows their strength vs.~their angular scale. It can be very well described by the six-parameter $\Lambda$CDM model, which takes into account dark energy as a cosmological constant $\Lambda$ and cold dark matter (CDM). Especially the third acoustic peak in the power spectrum is very sensitive to the total amount of dark matter in the Universe, see Fig.~\ref{fig::cmb}. The precise contributions of the different components to the energy density of the Universe as derived from the measurements of the Planck satellite are $\Omega_\Lambda = 0.6911(62)$, $\Omega_\tnl{CDM} = 0.2589(57)$ and $\Omega_\tnl{baryon} = 0.0486(10)$~\cite{Ade:2015xua}. The position of the first acoustic peak shows that the Universe's geometry is flat.

\subsection{Dark Matter Distribution in the Galaxy}
\label{sec::distribution}

Direct detection experiments search for events induced by particles from the Milky Way's dark matter halo. Their distribution and kinematics governs the expected signal and is thus essential for the interpretation of the experimental results. The exact distribution could not yet be measured, however, it can be modeled using the knowledge derived from the gravity-based observations discussed in Section~\ref{sec::evidence}.

The most simple model assumes that dark matter is a collisionless gas with an isotropic initial velocity distribution. Its equation-of-state relating pressure~$p$ and matter density~$\rho$ is given by
\begin{equation}
 p(r) = \rho(r) \ \langle (v-\overline{v})^2 \rangle = \rho(r) \ \sigma^2 \tn{,}
\end{equation}
with the velocity dispersion~$\sigma^2$. In hydrostatic equilibrium, the pressure of the dark matter gas balances the gravitational pull towards the center of the dark matter halo and the density profile can be calculated to be 
\begin{equation}\label{eq::iso}
 \rho(r)=\frac{\sigma^2}{2 \pi G \ r^2} \tn{.}
\end{equation}
The $r^{-2}$-dependence  of this ``isothermal sphere'' correctly leads to the observed flat galactic rotation profiles (see Section~\ref{sec::evidence}). 

\paragraph{Standard Halo Model} The Standard Halo Model commonly used for the interpretation of direct detection experiments assumes that the dark matter particles are distributed in an isotropic isothermal sphere with a Maxwellian velocity distribution
\begin{equation}
 f(\vec{v}) = N \ \exp \left( -\frac{3 | \vec{v} |^2}{2 \sigma^2} \right) \tn{,}
\end{equation}
which follows from the solution of the Boltzmann equation for collisionless particles, with the normalization $N \propto v^2$. In contrast to the stellar galactic disk, the halo does not rotate. The velocity dispersion~$\sigma(R)$ is related to the average circular velocity $v_c(R)$ at which objects at a distance $R$ orbit around the galactic center: $\sigma(R) = \sqrt{3/2} \ v_c(R)$. The local circular velocity at the solar distance $R_0\simeq8.0$\,kpc is $v_c(R_0)= 220$\,km/s. (Note that the Sun has a peculiar motion with respect to $v_c$.) Formally, the velocity distribution $f(\vec{v})$ is infinite, however, particles with velocities above the escape velocity $v_\tnl{esc} = \sqrt{2 \phi}$ are not bound by the gravitational potential~$\phi$ of the galaxy and $f(\vec{v})$ has to be truncated. By analyzing samples of stars with the highest measured velocities one obtains $v_\tnl{esc}\simeq544$\,km/s, with a 90\% confidence range from $498-608$\,km/s~\cite{Smith:2006ym}. 

The question whether the assumed Maxwell-Boltzmann velocity distribution of the Standard Halo Model is an over-simplification can be studied with large $N$-body simulations (for a review, see~\cite{Kuhlen:2012ft}). Starting from initial conditions given by the CMB fluctuations, these simulations model the evolution of parts of the Universe taking into account only dark matter ($=$only gravitational interactions) or dark matter with baryonic feedback (from supernovae and radiative feedback from massive stars) and provide information on Milky Way-like dark matter halos. Recent studies indicate that velocity distributions taken directly from high-resolution hydrodynamical simulations lead to direct detection results that are very similar to the ones based on the Standard Halo Model~\cite{ref::bozorgnia2016}. Another important conclusion from simulations is that the dark matter distribution at the solar distance $R_0$ is expected to be rather smooth, without relevant substructure. However, it is worth mentioning that some recent observations, such as the distribution and motion of satellite dwarf galaxies, are in conflict with simulations which predict them to be uniformly distributed and moving in random directions~\cite{ref::mueller2018}. 

\paragraph{Dark Matter Density} The simple isothermal sphere, Equation~(\ref{eq::iso}), shows a divergence at the galactic center. The precise shape of the density profile for $r \to 0$ is unknown and different possibilities were derived from $N$-body simulations. Some solutions exhibit a flat core, others are more cuspy. Recent studies of dwarf galaxies indicate that the ``cuspyness'' of the density profile depends on the star formation rate which drives fluctuations in the gravitational potential: galaxies with longer lasting star formation have more shallow dark matter cores~\cite{Read:2018fxs}. 

Nevertheless, at the solar radius $R_0$, all models basically agree, such that the knowledge of the density profile in the Milky Way center is irrelevant for direct detection experiment. (It is, however, crucial for the interpretation of results from indirect detection searches looking for particle excesses due to dark matter annihilation in the galactic center; for a review, see~\cite{Gaskins:2016cha}.) The local dark matter density $\rho_0$ can be derived from the measured rotation curve of the Milky Way (assuming spherical symmetry) or from the vertical kinematics and position of stars in the solar neighborhood. The canonical value adopted for the interpretation of direct detection experiments is $\rho_0=0.3$\,GeV/$c^2$/cm$^3$. However, the number has a rather large uncertainty of $\sim$50\% as individual measurements show considerable variations~\cite{Read:2014qva}. A recent measurement (2017) using Sloan Digital Sky Survey data yields $\rho_0=0.46 {+0.07 \atop -0.09}$\,GeV/$c^2$/cm$^3$~\cite{Sivertsson:2017rkp}. In the absence of a signal, it is more important for the direct detection community to adopt one common value to compare the different experiments rather than adopting an updated result: a change of $\rho_0$ will simply shift an exclusion limit/detection claim up or down in cross section by the same amount.

\subsection{Weakly Interacting Massive Particles (WIMPs)}
\label{sec::wimps}

Dark matter forms gravitationally interacting structures and it is generally assumed that it is of particle nature, similar to all ``normal'' matter. Dark matter in the form of Massive Compact Halo Objects (MACHOs), e.g., brown dwarfs, lonely planets, black holes populating the galactic halo, was not found in microlensing surveys in numbers sufficient to explain the required amount of dark matter~\cite{Tisserand:2006zx,Brandt:2016aco}. Primordial black holes (PBHs) produced in the very early phase of the Universe before big bang nucleosynthesis (and thus not being part of the 5\%~budget of baryonic matter in the Universe) came again into the focus of interest after the first observations of gravitational waves from the inspiral event of black holes with unexpected high masses of 20-30\,M$_\odot$~\cite{ref::ligo}. However, the total number of detected events is too small to explain the full amount of dark matter; PBHs in this mass range could only constitute $\sim$1\%~\cite{ref::pbh1}. Other constraints come from microlensing searches of stars~\cite{ref::pbh2,ref::pbh3}, type~Ia supernovae~\cite{ref::pbh4}, or cosmic ray data from Voyager~I~\cite{Boudaud:2018hqb} and basically exclude PBHs as being all of the dark matter in any mass range. The recent observation of gravitational waves in coincidence with electromagnetic radiation~\cite{ref::ligo2017} also placed tight constraints on the propagation speed of gravitational waves and thus allows ruling out various popular models which aim at explaining the indications for the existence of dark matter, most notably the flat rotation curves (see Sect.~\ref{sec::evidence}), by modifying the laws of gravity.

Dark matter made from baryons can be excluded by measurements of the primordial abundance of light elements produced in the big bang nucleosynthesis ($^2$H, $^3$He, $^4$He, $^7$Li, for a recent review see~\cite{Fields:2011zzb}) and by precision studies of the CMB power spectrum, which constrains the baryonic matter content of the Universe to $\Omega_b \approx 5$\%~\cite{Ade:2015xua}. As the Standard Model of Particle Physics does not contain a single suitable dark matter candidate it is assumed that dark matter must be made of one (or more?) new particle(s) which has not been directly detected yet. (The free-streaming length of the massive but very light-weight neutrinos would wash out the observed large-scale structure of galaxies in the Universe).    

\begin{figure}[b!]
\centering
\includegraphics[width=0.8\textwidth]{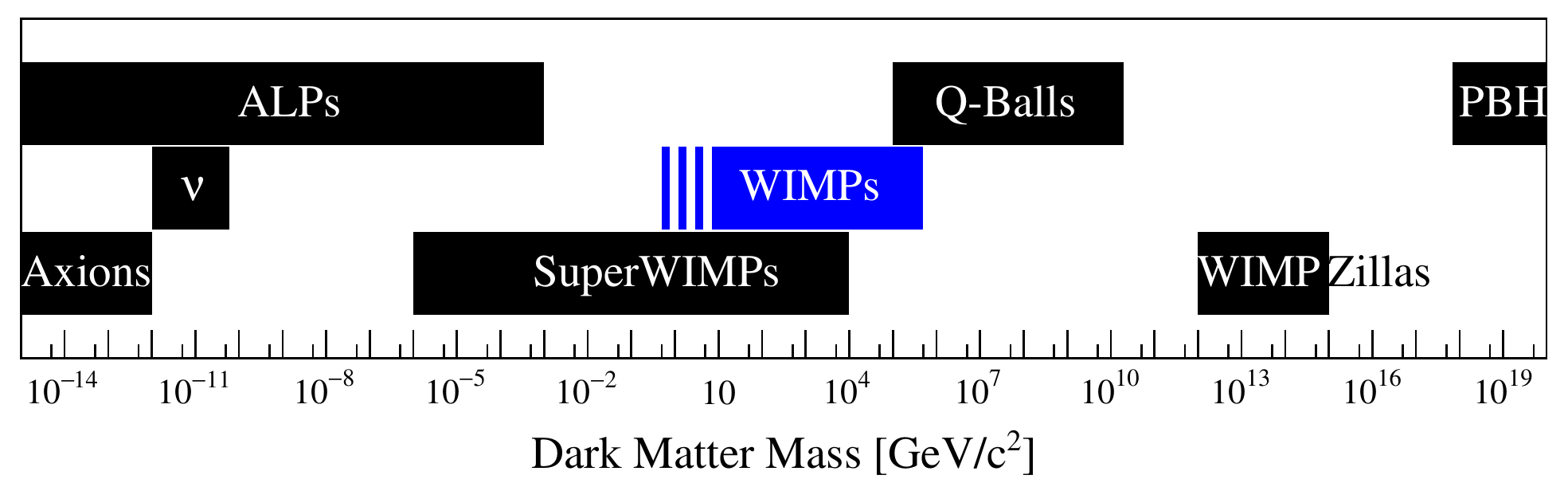}
\caption{The predicted mass range for the various prominent dark matter particle candidates covers many orders of magnitude. This review focuses on the Weakly Interacting Massive Particle (WIMP); the low-mass region discussed later is indicated by the dashed bar. } \label{fig::candidates}
\end{figure}

There are many suggestions for ``new physics'' particles that would solve the dark matter puzzle, see Fig.~\ref{fig::candidates}. One important class of potential dark matter candidates are Weakly Interacting Massive Particles (WIMPs). They arise naturally in various theories beyond the Standard Model, e.g., as the lightest supersymmetric particle in supersymmetric theories (LSP; in many models this is the neutralino~$\chi$)~\cite{ref::nilles, ref::lahanas, ref::peskin}, or as the lightest Kaluza-Klein particle (LKP) in theories with extra spacetime dimensions~\cite{ref::hooper}. The lightest particle in Little Higgs models is also a possible WIMP candidate~\cite{ref::chen2014} if -- as required for almost every theory -- there is some new quantum number (``parity'') which is conserved and prevents the decay of the WIMP candidate into lighter Standard Model particles. 
It is important to stress that none of these models was proposed to explain the dark matter problem -- the dark matter candidate comes for free.

WIMPs constitute a rather model-independent generic class of dark matter candidates, with masses in the 1~to $10^5$\,GeV/$c^2$ range and interaction cross sections from $10^{-41}$ to $10^{-51}$\,cm$^2$~\cite{ref::roszkowski2004}. The well-studied supersymmetric models usually predict WIMPs with masses around 100\,GeV/$c^2$, however, other models favor different mass ranges. Asymmetric Dark Matter~\cite{Zurek:2013wia}, for example, relates the observed matter-antimatter asymmetry in the Universe with dark matter and predicts 
\begin{equation}
m_\chi \approx 5 \ m_\mathrm{proton} \approx 5 \ \tn{GeV}/c^2 \tn{.}   
\end{equation}

The massive and electrically neutral WIMPs are assumed to be thermally produced in the early universe and are now moving with non-relativistic velocities, which makes them the prime candidate for ``cold'' dark matter. (A summary of dark matter production can be found, e.g., in~\cite{Baer:2014eja}.) The interesting observation that the thermal production of WIMPs with weak-scale cross sections naturally leads to the correct relic dark matter abundance of $\Omega_\tnl{CDM} \approx 0.25$ after their freeze-out from the thermal plasma is usually referred to as the ``WIMP miracle''. The result is almost independent of the WIMP mass and serves to further motivate the case for WIMPs. 

While many alternative dark matter candidates exist -- prominent examples are heavy right-handed neutrinos~\cite{Shaposhnikov:2006xi} and axions or axions-like particles (ALPs)~\cite{Ringwald:2012hr}, see Fig.~\ref{fig::candidates} -- the WIMP still is the prime target for experimental searches. The remaining part of this article will thus concentrate on the WIMP, which, despite of many years of effort, was not yet discovered. However, many detectors with an improved sensitivity will get online in the near future.

\section{Direct Detection Basics}
\label{sec::ddbasics}

Information about the dark matter particle can be inferred indirectly from astrophysical and cosmological observations. Examples are limits on its self-interaction strength from colliding galaxy clusters~\cite{Harvey:1462}, such as the Bullet cluster~\cite{Clowe:2006eq}, or constraints from large scale structures~\cite{Frenk:2012ph} and from the Universe's ionization history~\cite{Lopez-Honorez:2017csg}. 

Three more direct approaches are pursued in order to eventually measure the dark matter particle physics properties, such as mass, coupling and interaction cross section, with baryonic matter: these are the \emph{production} of dark matter particles at hadron colliders such as the LHC~\cite{Kahlhoefer:2017dnp}, the detection of the decay products following dark matter \emph{annihilation} processes in regions of high dark matter densities (e.g., the galactic center, the Sun, dwarf galaxies)~\cite{Gaskins:2016cha}, and the direct detection of WIMP-nucleus \emph{scattering} processes in ultra-sensitive low-background experiments. In this section, we will review the physics of direct detection and discuss the characteristics of the used detectors.

\subsection{Rates, Spectra and Interactions}
\label{sec::rates}

The possibility to directly detect dark matter particles in the form of WIMPs was first discussed by Goodmann and Witten~\cite{ref::goodmannwitten}. Since the WIMP carries no electric charge, in most scenarios it will not interact with the atomic electrons but will instead elastically scatter off the atomic nucleus. The momentum transfer gives rise to a nuclear recoil which might be detectable. The discussion below largely follows~\cite{Lewin:1995rx}.

\paragraph{Rate and Number of Events}
The expected rate of WIMPs scattering off a target nucleus of mass $m_N$ is given by
\begin{equation}\label{eq::rate}
\frac{dR}{dE_\tnl{nr}} = \frac{\rho_0 M}{m_N m_\chi} \int_{v_\tn{\tiny min}}^{v_\tn{\tiny esc}} v f(v) \frac{d \sigma}{dE_\tnl{nr}} \ dv\textnormal{.}
\end{equation}
$E_\tnl{nr}$ is the nuclear recoil (NR) energy, $m_\chi$ denotes the WIMP mass, $\sigma$ is the scattering cross section. $M$ is the target mass of the detector. The astrophysical parameters describe the WIMP distribution in the Milky Way (see also Section~\ref{sec::distribution}): $f(v)$ is the normalized WIMP velocity distribution and $\rho_{0}=0.3$\,GeV/$c^{2}$/cm$^3$ the local dark matter density. All velocities are defined in the detector's reference frame. The minimal velocity required for a WIMP to induce a nuclear recoil of energy $E_\tnl{nr}$ is
\begin{equation}\label{eq::mu}
v_\tnl{min}=\sqrt{\frac{E_\tnl{nr} m_N}{2} \frac{(m_N+m_\chi)^2}{(m_N m_\chi)^2}} = \sqrt{\frac{E_\tnl{nr} m_N}{2} \frac{1}{\mu^2}}                                                                                                                                                                                                                        \textnormal{.}
\end{equation}
WIMPs with a velocity above the escape velocity $v_\tnl{esc}=544$\,km/s~\cite{Smith:2006ym} will not be bound to the potential well of the Milky Way. $\mu$~denotes the reduced mass of the nucleon-WIMP system.

\begin{figure}[t]
\centering
\includegraphics[width=0.7\textwidth]{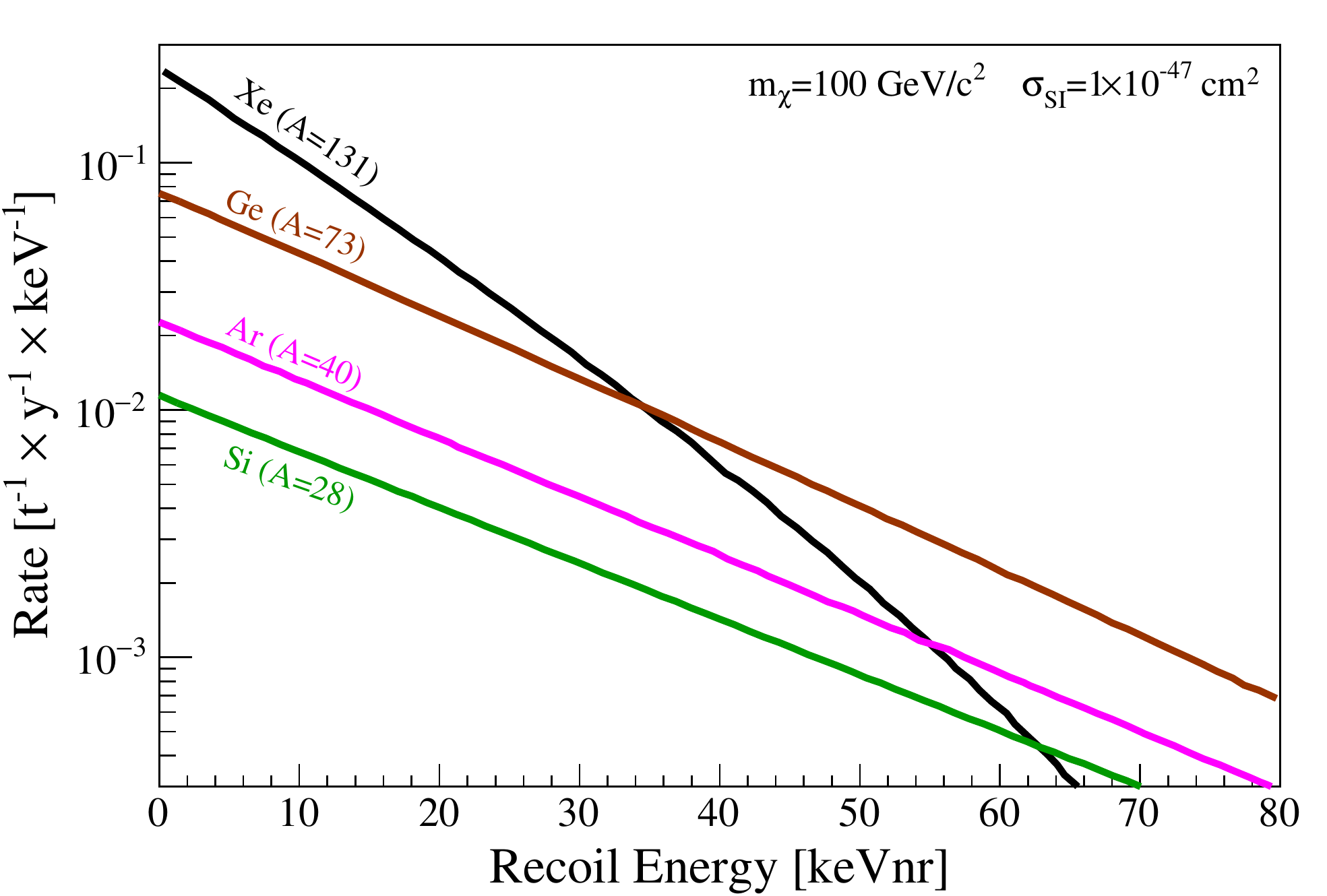}
\caption{Nuclear recoil spectra induced by a $m_\chi=100$\,GeV/$c^2$ WIMP in various common target materials, assuming a spin-independent WIMP-nucleon cross section $\sigma_\tnl{SI}=1 \times 10^{-47}$\,cm$^2$. The rate for spin-independent interactions increases with $A^2$, i.e., prefers heavy target materials. On the other hand, the rate for large nuclei (Xe, I, W) decreases at higher energies due to the form factor suppression.} \label{fig::wimpspectra}
\end{figure}

The observed number of events in an experiment running for a time $T$ is obtained by integrating Eq.~(\ref{eq::rate}) from the threshold energy $E_\tnl{low}$ to the upper boundary  $E_\tnl{high}$:
\begin{equation}
N=T \int_{E_\tn{\tiny low}}^{E_\tn{\tiny high}} dE_\tnl{nr} \ \epsilon(E_\tnl{nr}) \ \frac{dR}{dE_\tnl{nr}} \textnormal{,} 
\end{equation}
with the (typically energy-dependent) detector efficiency~$\epsilon$. The maximum recoil energy $E_\tnl{high}=2 \mu^2  v_\tnl{esc}^2 m_N^{-1}$ is given by kinematics and much less relevant than the energy threshold $E_\tnl{low}$ as the differential rate, Eq.~(\ref{eq::rate}), is eventually described by a simple steeply falling exponential function~\cite{Lewin:1995rx}  (see Fig.~\ref{fig::wimpspectra})
\begin{equation}
\frac{dR}{dE_\tnl{nr}} \propto \exp \left(- \frac{E_\tnl{nr}}{E_0} \frac{4 m_\chi m_N}{(m_\chi + m_N)^2} \right) \tn{,}
\end{equation}
with $E_0$ being the most probable kinetic energy of the incident WIMP and typical recoil energies of ${\cal O}$(10)\,keV$_\tnl{nr}$ only. The nuclear recoil energy is given in keV$_\tnl{nr}$ (nuclear recoil equivalent), which is different from the electronic recoil scale (keV$_\tnl{ee}$) due to quenching effects caused by the different energy-loss mechanisms (see Section~\ref{sec::generation}).

\paragraph{Annual Modulation and Directionality} The relevant velocity for the WIMP-nucleon scattering in an Earth-based detector is not only the local circular velocity $v_c=220$\,km/s (as discussed in Section~\ref{sec::distribution}), but it is the one of the Earth moving through the dark matter halo with
\begin{equation}
 v_E = v_\odot + v_\oplus \cos(\theta) \cos \left[ \omega (t-t_0) \right] \tn{,}
\end{equation}
where $v_\odot = v_c + 12$\,km/s describes the motion of the Sun with respect to $v_c$ and $v_\oplus=30$\,km/s is the speed of the Earth orbiting around the Sun. $\theta \approx 60^\circ$ is the inclination angle measured between the Earth's orbit and the galactic plane. The angular frequency $\omega = 2 \pi/T$ is defined by $T=1$\,y and the phase is fixed to $t_0=$\,June 2, when $v_\odot$ and $v_\oplus$ add up maximally. This detail leads to two observational consequences, which both are used in experiments.

\begin{figure}[t]
\centering
\includegraphics[width=0.7\textwidth]{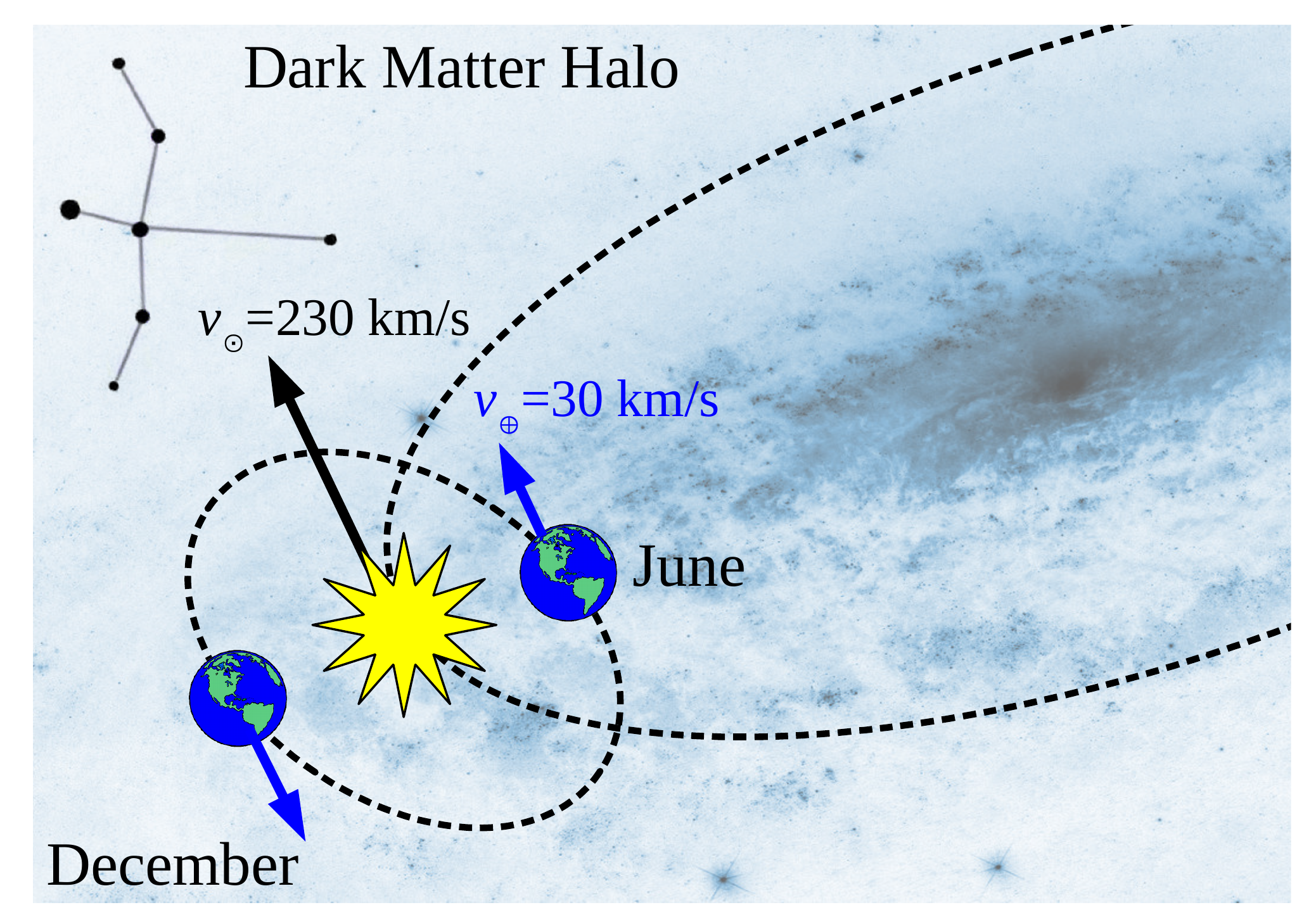}
\caption{Illustration of the Sun-Earth system moving around the galactic center and through the dark matter halo in the direction of the constellation Cygnus. The varying vector addition of the velocities over the course of a year is expected to induce a modulating dark matter signature.  } \label{fig::modulation}
\end{figure}

The different mean incident WIMP velocity in summer ($v_\odot$ and $v_\oplus$ parallel) and in winter ($v_\odot$ and $v_\oplus$ anti-parallel) leads to a harder or softer WIMP spectrum, respectively, see Fig.~\ref{fig::modulation}. Given a constant detector threshold, the number of observed dark matter signal events $S(t)$ will thus \emph{modulate} over the course of the year~\cite{Drukier:1986tm}:
\begin{equation}
 S(t)=B + S_0 + S_m \cos \left[ \omega (t-t_0) \right] \tn{.}
\end{equation}
The modulated part of the signal~$S_m$ is rather small, ${\cal O}(v_\oplus/v_c) \sim 5\%$, and most of the signal is unmodulated~$S_0$. The same is expected for detector-related backgrounds $B$, which are typically much larger than $S_0$. However, it is known that some possible background sources also show a seasonal modulation, e.g., the atmospheric neutrino flux depends on the temperature-dependent density of the atmosphere~\cite{Bellini:2012te}. Modifications to the simple dark matter distribution model, e.g., dark matter streams etc., would significantly alter the expected signal and while astrophysical uncertainties of the local dark matter distribution have generally a rather small impact for ``standard'' direct detection experiments, they get more significant for such annual modulation searches~\cite{Green:2017odb}. A detection of dark matter based solely on the modulation signature is only possible in a statistical way, i.e., ${\cal O}(10^4)$ signal events are required.

An even stronger dark~matter-specific signature that can be used to reduce background events is to exploit the fact that $\vec{v}_c \approx \vec{v}_\odot$ points at a specific direction in the sky, towards the constellation of Cygnus. Thus the ``WIMP wind'' is expected to come from this region of the sky. The differential recoil rate depends on the angle~$\phi$ between the direction of the WIMP and the direction of the nuclear recoil measured in the frame of the galaxy,
\begin{equation}
 \frac{d^2R}{dE_\tnl{nr} \ d\phi} \propto \exp \left[ \frac{2(v_\odot \cos \phi - v_\tnl{min})^2}{3 v_\odot^2} \right] \tn{,}
\end{equation}
and is sharply peaked towards the Sun's direction of motion. The Earth's daily rotation thus constantly changes the signal direction observed in a detector. $v_\tnl{min}$ was defined in Eq.(\ref{eq::mu}). As most backgrounds are expected to be uniformly distributed, or come from the Sun as, e.g., solar neutrinos, a measurement of the \emph{track direction} (in addition to the recoil energy) could be used to distinguish a dark matter signal from background events~\cite{ref::spergel1988}. The additional directional information (of a large statistics sample) would allow the study of many particle and astrophysics dark matter properties. The experimental challenge is that the track length of keV-scale nuclear recoils is very short, $r$\,$<$\,1\,mm, and hard to reconstruct even in low-pressure gas detectors~\cite{Mayet:2016zxu} (see Section~\ref{sec::detectors}).

\paragraph{Spin-Independent and Spin-Dependent Interactions}
Because of its large de Broglie wavelength, the WIMP interacts coherently with all nucleons in the target nucleus. The WIMP-nucleus scattering cross section in Eq.~(\ref{eq::rate}) is velocity and recoil-energy dependent and given by
\begin{equation}
\frac{d\sigma}{dE_\tnl{nr}} = \frac{m_N}{2 v^2 \mu^2} \left( \sigma_{SI} F_{SI}^2(E_\tnl{nr}) + \sigma_{SD} F_{SD}^2(E_\tnl{nr}) \right)\textnormal{.}
\end{equation}
Since the interaction of WIMPs with baryonic matter is a priori unknown, the cross section consists of two terms for spin-independent (SI, which can be described by a scalar ${\cal L}_S \sim \overline{\chi}\chi \overline{q}q$ or vector ${\cal L}_V \sim \overline{\chi} \gamma^\mu \chi \ \overline{q} \gamma^\mu q$ effective 4-fermion Lagrangian) and spin-dependent (SD, axial-vector ${\cal L}_A \sim \overline{\chi} \gamma^\mu \gamma_5 \chi \ \overline{q} \gamma^\mu \gamma_5 q$) couplings. At small momentum transfers $q=\sqrt{2 m_n E_\tnl{nr}}$ all partial waves of the nucleons add up and the WIMP scatters coherently off the entire nucleus. For higher~$q$ the WIMP's de Broglie wavelength $\lambda=h/q$ is reduced and only part of the nucleus participates in the interaction. This loss of coherence is accounted for by the finite form factors $F_{SI}$ and $F_{SD}$, which are only relevant for heavy WIMP targets such as Xe, I or W (see Fig.~\ref{fig::wimpspectra}). A usual choice for the (spin-independent) form factor is the analytical approximation of Helm~\cite{ref::helm,Lewin:1995rx} using the dimensionless parameter $x=qR_N/\hbar$:
\begin{equation}
F(x) = \frac{3 \ j_1(x)}{x} \ \exp \left( -\frac{(xs)^2}{2 R_N^2} \right) \tn{.}
\end{equation}
$j_1(x)=\sin(x)/x^2-\cos(x)/x$ is the spherical Bessel function of the first kind. This expression is based on the Woods-Saxon potential describing the nucleus of atomic number $A$, with a radius $R_N \approx 1.2 \ A^{1/3}$\,fm and a skin-thickness $s \approx 0.5$\,fm. (In general, the parameters have to be precisely obtained for every nucleus.) 

The \emph{spin-independent}~(SI) cross section is given by
\begin{equation}\label{eq::si}
\sigma_{SI}=\sigma_n \frac{\mu^2}{\mu_n^2}\frac{(f_pZ+f_n(A-Z))^2}{f_n^2} = \sigma_n \frac{\mu^2}{\mu_n^2} A^2 \textnormal{.}
\end{equation}
$\mu$ is the WIMP-nucleus reduced mass as defined in Eq.~(\ref{eq::mu}), and $\mu_n$ is the reduced mass of the WIMP-nucleon system. To allow for the comparison of different target nuclei employed in different experiments, the WIMP-nucleus cross section $\sigma$ is usually converted to a WIMP-nucleon cross section $\sigma_n$. $f_p$ and $f_n$ describe the WIMP coupling strength to protons and neutrons, respectively. The second expression assumes $f_p=f_n$ which eventually leads to an $A^2$ dependence of the cross section: This implies that heavier target nuclei are expected observe higher event rates, however, at the same time the expected recoil energy $E_\tnl{nr}$ is smaller which poses bigger challenges on the detector's thresholds. 

The differential cross section for \emph{spin-dependent}~(SD) interactions, where the WIMP is assumed to be a (Dirac or Majorana) fermion coupling to unpaired nuclear spins~$J$, reads~\cite{ref::engel}
\begin{equation}\label{eq::sd}
\frac{d\sigma_{SD}}{\textnormal{d}|\vec{q}|^2} = \frac{8 G_F^2}{\pi v^2} \left[ a_p \langle S_p \rangle + a_n \langle S_n \rangle \right]^2 \frac{J+1}{J} \frac{S(|\vec{q}|)}{S(0)} \textnormal{,} 
\end{equation}
with the momentum transfer $\vec{q}$. $\langle S_{p} \rangle$ and $\langle S_{n} \rangle$ are the expectation values of the total spin operators for protons and neutrons in the the target nucleus which have to be calculated and show some model dependence~\cite{Toivanen:2009zza,Menendez:2012tm}. There is no $A^2$-scaling from coherence effects as in Eq.~(\ref{eq::si}), instead the cross section depends on the total nuclear spin $J$ of the target nucleus as well as its spin-structure function $S(|\vec{q}|)$. While heavy nuclei are typically more sensitive to SI-interactions, the situation is different for SD-scattering, as only the nucleus' spin structure is relevant: nuclei without unpaired spins, e.g., argon targets which consist of $_{18}^{40}$Ar, $_{18}^{38}$Ar and $_{18}^{36}$Ar, are ``blind'' while the nuclear spin-structure of the rather light nucleus $^{19}_{\ 9}$F makes it an excellent target to probe SD-interactions. Neutrons and protons typically contribute differently to the total spin of the target such that the SD-results are commonly quoted assuming that WIMPs couple either only to neutrons (coupling strength $a_p=0$) or to protons ($a_n=0$): the above-mentioned $^{19}_{\ 9}$F is very sensitive to spin-dependent WIMP-proton scattering while $^{73}_{32}$Ge, $^{129}_{\ 54}$Xe and $^{131}_{\ 54}$Xe are sensitive to spin-dependent WIMP-neutron interactions. The expression $[ a_p \langle S_p \rangle + a_n \langle S_n \rangle ]^2 (J+1)/J$ in Eq.~(\ref{eq::sd}) allows defining an isotope-specific sensitivity factor for SD interactions. Its relative strength for various isotopes used as WIMP targets is given in Table~\ref{tab::sd}, assuming a coupling to the unpaired nucleon only. Chiral two-body currents lead also to a sensitivity to SD-interactions with an even number of nucleons, however at a reduced sensitivity~\cite{Klos:2013rwa}. 

\begin{table}
\caption{Relative strength of different isotopes used as targets in direct detection experiments for spin-dependent WIMP searches. It is normalized the $^{19}$F, the target with the highest sensitivity to spin-dependent coupling to protons. The expectation values $\langle S_{p,n} \rangle$ to calculate the relative strength are taken from~\cite{Klos:2013rwa}. }\label{tab::sd}
\centering
\small
\begin{tabular}{crrccr}
\hline \hline
Isotope & $Z$ & Abundance & Spin & Unpaired Nucleon & Relative Strength \\ \hline
$^{19}$F \ & 9 & 100.0\% & 1/2  & proton  & 100 \\
$^{23}$Na & 11 & 100.0\% & 3/2 & proton &  12 \\
$^{27}$Al & 13 & 100.0\% & 5/2 & proton &  22 \\
$^{29}$Si & 14 & 4.7\% & 1/2 & neutron &  11 \\
$^{73}$Ge & 32 & 7.7\% & 9/2 & neutron &  34 \\
$^{127}$I \ & 53 & 100.0\% & 5/2 & proton &  24 \\
$^{129}$Xe & 54 & 26.4\% & 1/2 & neutron & 47 \\
$^{131}$Xe & 54 & 21.2\% & 3/2 & neutron & 18 \\ \hline \hline
\end{tabular}
\end{table}

The ``traditional'' analyses in terms of SI and SD WIMP-nucleon ($\chi$-$N$) interactions are a subset of more general \emph{effective field theories} (EFTs)~\cite{Fan:2010gt,Fitzpatrick:2012ix,Hoferichter:2016nvd} which construct a large number of relativistic and non-relativistic operators to describe the various possible 4-point-interactions, assuming a heavy mediator particle. EFTs provide a possibility to directly compare results from direct detection and collider searches, however, the heavy mediator assumption has important consequences for the interpretation of LHC searches as one needs to ensure that the kinematics of the WIMP production process is described accurately~\cite{Buchmueller:2013dya}. This problem is overcome by more complex \emph{simplified models} which replace the 4-point-interaction by $s$- or $t$-channel exchange of the mediator. The SI and SD interactions are mainly described by the non-relativistic EFT operators ${\cal O}_1 = 1_\chi 1_N$ and ${\cal O}_4=\vec{S}_\chi \cdot \vec{S}_N$, respectively, however, the experimental result of a WIMP search can be interpreted in a plethora of operators~\cite{Aprile:2017aas}; some of these would produce very different recoil spectra than expected for the ``standard'' SI case. 

The shape of the expected recoil spectrum is determined by the interaction model and kinematics (e.g., $f(v)$, $m_N$, $m_\chi$ for SI scattering). If the local dark matter density~$\rho_0$ is fixed, the expected total rate depends on the scattering cross section. SI rates of $\gtrsim$3\,events/ton/year have already been excluded which corresponds to WIMP-nucleon cross sections $\sigma_n\gtrsim 4 \times 10^{-47}$\,cm$^2$ for $m_\chi \sim 30$\,GeV/$c^2$~\cite{Aprile:2018dbl}. In order to be sensitive to such small SI cross sections, the optimal WIMP detector should have a large total mass $M$, a high mass number $A$, a low energy threshold $E_{low}$, an ultra-low background and the ability to distinguish between (nuclear recoil) signal and (electronic recoil) background events. To focus on a special part of the parameter space or different interaction models, as well as practical reasons might require compromises or departures from these general rules: e.g., very light targets are favorable for SI interactions of low-mass WIMPs. Various detector designs are discussed in the next Section.

\subsection{Signal Generation in WIMP Detectors}
\label{sec::generation}

The total energy loss of a recoil in a WIMP detector can be described by
\begin{equation}
\left(\frac{dE}{dx} \right)_\tnl{tot} = \left(\frac{dE}{dx} \right)_\tnl{elec} + \left(\frac{dE}{dx} \right)_\tnl{nucl} \textnormal{.}
\end{equation}
The majority of the nuclear recoil energy induced by a WIMP-nucleus interaction is typically lost to heat, i.e., to nuclear energy losses  $\left(\frac{dE} {dx}\right)_\tnl{nucl}$ leading to atomic motion. In solid materials this gives rise to phonons. Smaller amounts of electronic energy losses are available to excite or ionize the target atoms. In some materials, the atomic excitation leads to scintillation light of a narrow emission spectrum, which can be detected by means of sensitive photosensors. Because of the smallness of the signal, the number of physical quanta (phonons, electrons, photons) available for detection is generally rather low. 

\begin{figure}[t!]
\centering
\includegraphics*[width=0.495\textwidth]{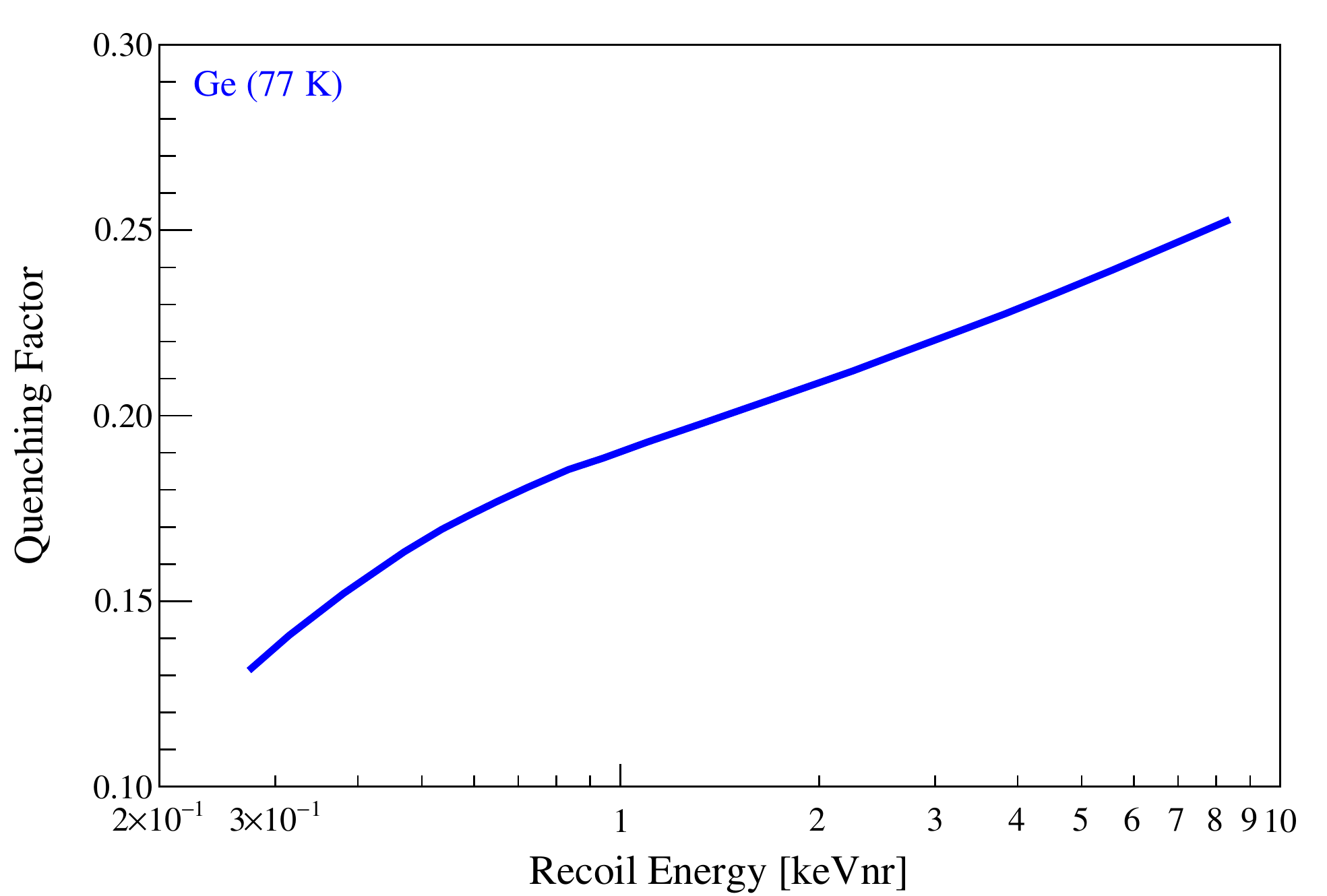}   
\includegraphics*[width=0.495\textwidth]{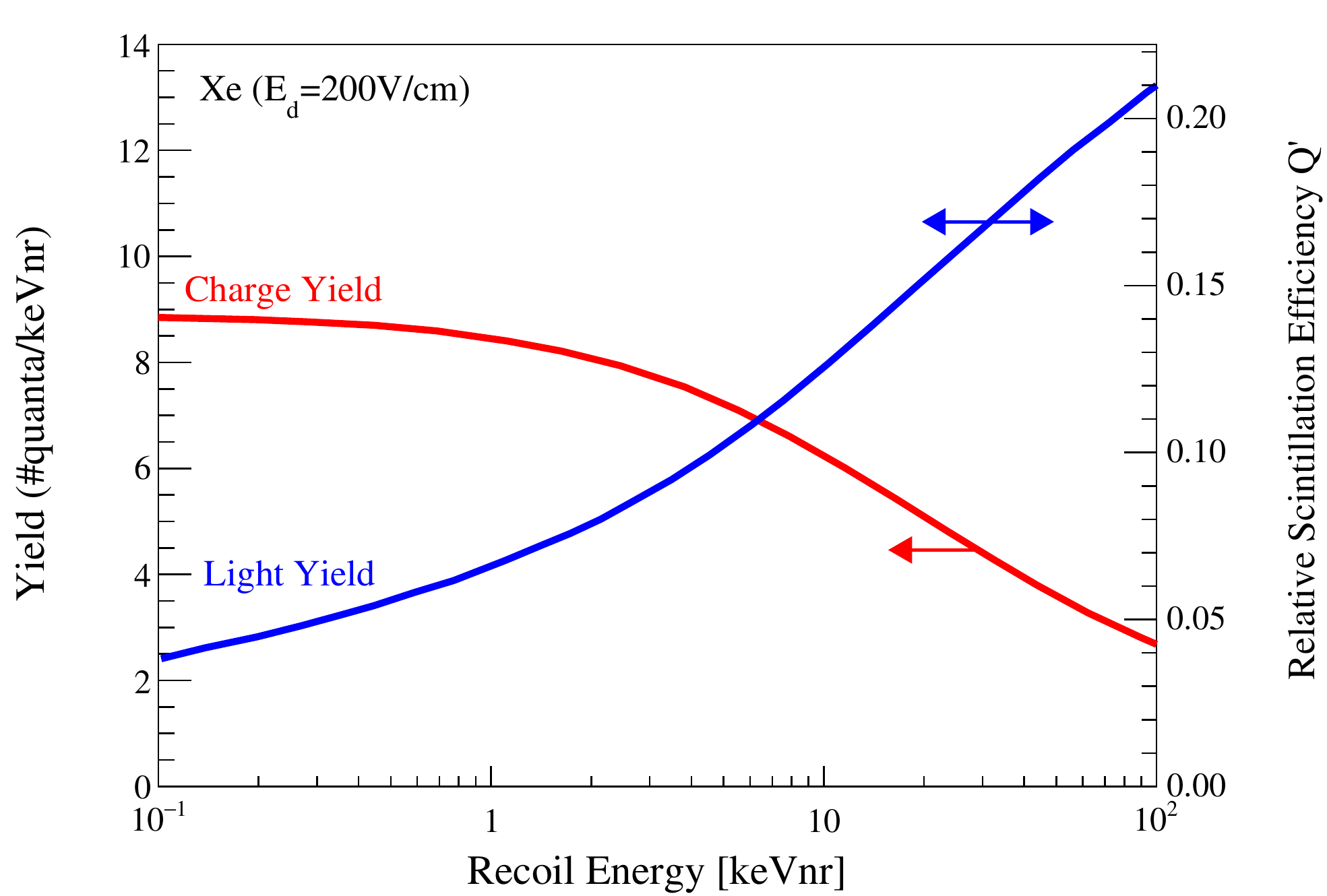} 
\caption{Examples for quenching factors. {\bf (Left)} The quenching factor describes by how much the ionization signal from an NR energy deposition in germanium at 77\,K is reduced compared to an ER of the same energy. Shown is a best-fit Lindhard model to experimental data~\cite{Scholz:2016qos}. {\bf (Right)} Light and charge yield (in number of photons or electrons per keV$_\tnl{nr}$) in a liquid xenon (LXe) TPC operated with a drift field of $E_d=200$\,V/cm. The light yield can also be expressed via a relative quenching factor (right axis, often referred to as ${\cal L}_\tnl{eff}$), here normalized to the light yield of a 31.2\,keV ER signal at $E_d=0$\,V/cm. Data from~\cite{ref::nest}. In both cases, the data exhibit uncertainties which are not shown: around 1\,keV$_\tnl{nr}$, the individual measurements have relative uncertainties around $\pm$1\% and $\pm$20-30\% for germanium and xenon, respectively. The systematic differences between measurements can be even larger.}
\label{fig::quenching}
\end{figure}

\paragraph{Signal Quenching} An experimental complication comes from the fact that the expected nuclear recoil signal of energy $E_\tnl{nr}$ is smaller than the same amount of energy deposited by an electronic recoil signal $E_\tnl{ee}$. This phenomenon is called \emph{signal quenching} and is due to the different energy-loss mechanisms of the two recoil types. In case of low-energy nuclear recoils, significantly more energy is lost to atomic motion (heat), which is often not detected. This leads to considerably smaller scintillation and ionization signals and makes their detection even more challenging. Signal quenching is typically energy dependent and has to be measured accurately in order to establish an energy scale for the detector. It is an intrinsic feature of the detection material and independent from the actually used detector (if detector-specific effects such as signal collection efficiency and thresholds are properly accounted for); this allows for quenching factor measurements in detectors specifically designed for this purpose. One usually distinguishes between absolute quenching factors $Q(E_\tnl{nr})$, which directly compare the reconstructed energy scales for electronic ($E_\tnl{ee}$, measured in electronic recoil equivalent energy, keV$_\tnl{ee}$) and nuclear recoils ($E_\tnl{nr}$, measured in nuclear recoil equivalent energy, keV$_\tnl{nr}$) in an energy-dependent fashion,
\begin{equation}
E_\tnl{ee}[\textrm{keV}_\tnl{ee}]=Q(E_\tnl{nr}) \times E_\tnl{nr}[\textrm{keV}_\tnl{nr}] \tn{,}
\end{equation}
and relative quenching factors $Q'(E_\tnl{nr})$, see Fig.~\ref{fig::quenching}. The latter describe the observed (light or charge) signal from a nuclear recoil interaction relative to a conveniently chosen electronic recoil calibration standard, e.g., the $E_\tnl{ee}=122$\,keV $\gamma$-line from $^{57}$Co or the 31\,keV conversion electron line from $^{83m}$Kr:
\begin{equation}
Q'(E_\tnl{nr}) = \frac{S(E_\tnl{nr})}{S(E_\tnl{ee}=\tn{fixed})} \tn{.} 
\end{equation}
An important example for a relative quenching factor is the relative scintillation efficiency ${\cal L}_\tnl{eff}$ of liquid noble gases (Ar, Xe): it describes how a scintillation light signal from a nuclear recoil energy deposition $E_\tnl{nr}$ in the liquid is reduced compared to a calibration standard (e.g., the 122\,keV line). 

For some materials such as Ge or Si~crystals, signal quenching can be reliably calculated using the Lindhard theory~\cite{ref::lindhard} which employs some basic assumptions on the motion of the recoiling nucleus in the material (e.g., mainly elastic nuclear collisions at low $E$). For other materials, e.g., Xe and Ar, the results are much worse and semi-analytical approaches are used instead~\cite{Szydagis:2011tk,ref::nest}, see Fig.~\ref{fig::quenching}.

Detectors which have the capability to actually measure the heat deposited in the target (see cryogenic detectors below) do not suffer from this complication; the heat measurement allows for a straightforward definition of the energy scale. The calibration can be done using artificially induced signals from a heater.

\subsection{Detectors for WIMP Searches}
\label{sec::detectors}

Several detector designs exploiting various target materials exist to detect one or two of the signal channels. (To date, no attempt to measure all three channels simultaneously in one detector was successful.) They can be put into the following categories:

\paragraph{A. Anorganic Crystal Detectors} The very first experimental attempt to directly detect WIMPs was realized shortly after the Goodmann and Witten suggestion~\cite{ref::goodmannwitten}. The experiment~\cite{Ahlen:1987mn} used a 0.72\,kg high purity Germanium (HPGe) crystal to search for dark matter-induced charge signals. Only a very small amount of energy is required to create and electron-hole pair in such semiconductor detectors (Ge: 2.9\,eV, Si: 3.6\,eV) which leads to many signal carriers and thus a very good energy resolution. However, the signals are rather slow ($\tau \sim 1$\,$\mu$s) and the large capacitance of Ge~diodes beyond the ${\cal O}$(1)\,kg-scale, which leads to high electronic noise levels, does not allow to build very massive detectors. Current state-of-the-art experiments use kg-scale p-type point contact HPGe crystals, which are able to achieve very low thresholds down to $\sim$160\,eV$_\tnl{ee}$~\cite{Jiang:2018pic}. Background events from the large n$^+$ surface can be reduced based on their longer rise times compared to bulk events, which is caused by the weaker electric fields next to the n$^+$ region~\cite{Aalseth:2010vx}. Some projects use Si-crystals ($Z=14$) in the form of CCD-chips to optimize their sensitivity to WIMPs at lower masses~\cite{Aguilar-Arevalo:2016ndq,Crisler:2018gci}. 

\begin{figure}[t!]
\centering
\includegraphics*[width=0.495\textwidth]{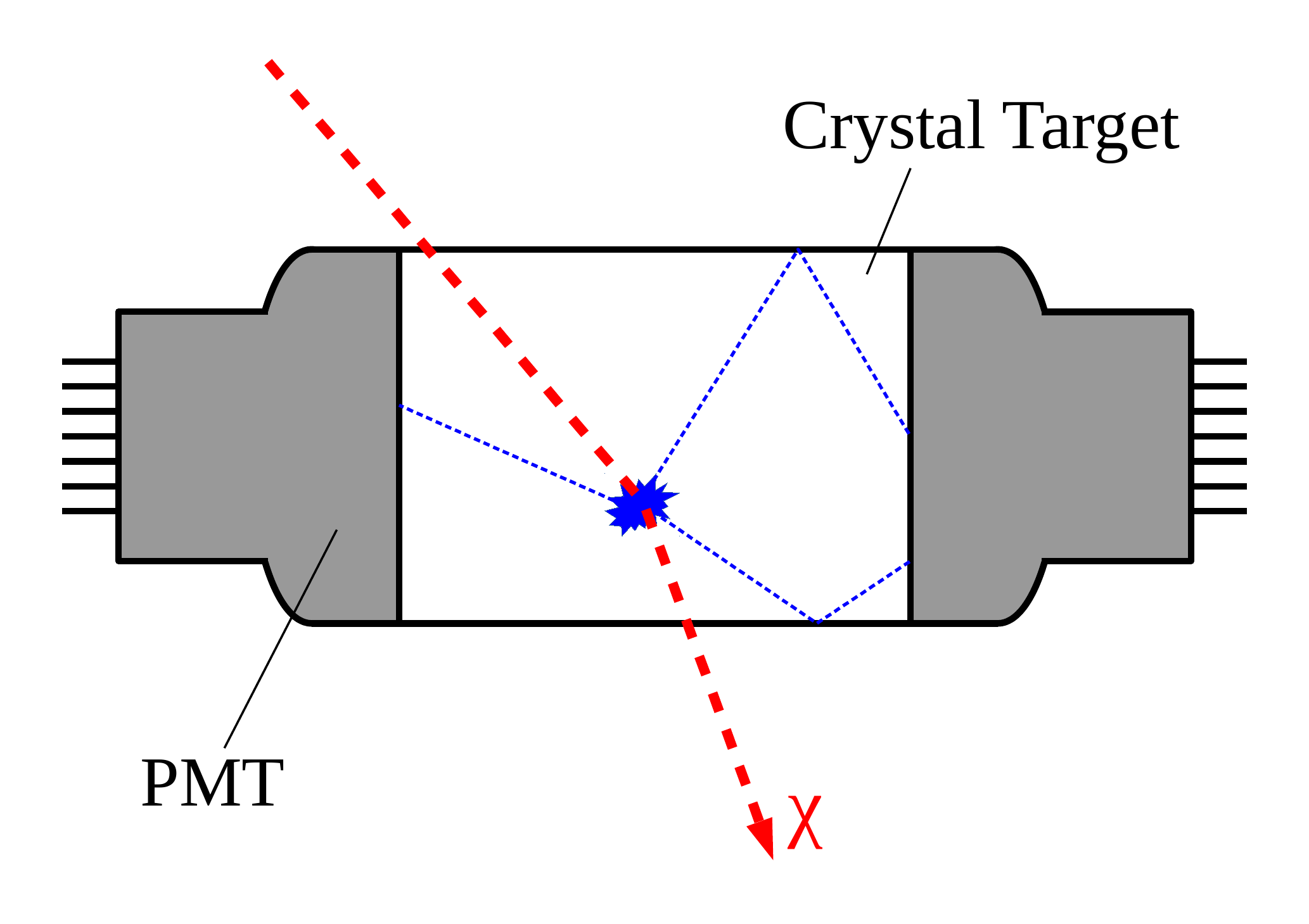}   
\includegraphics*[width=0.495\textwidth]{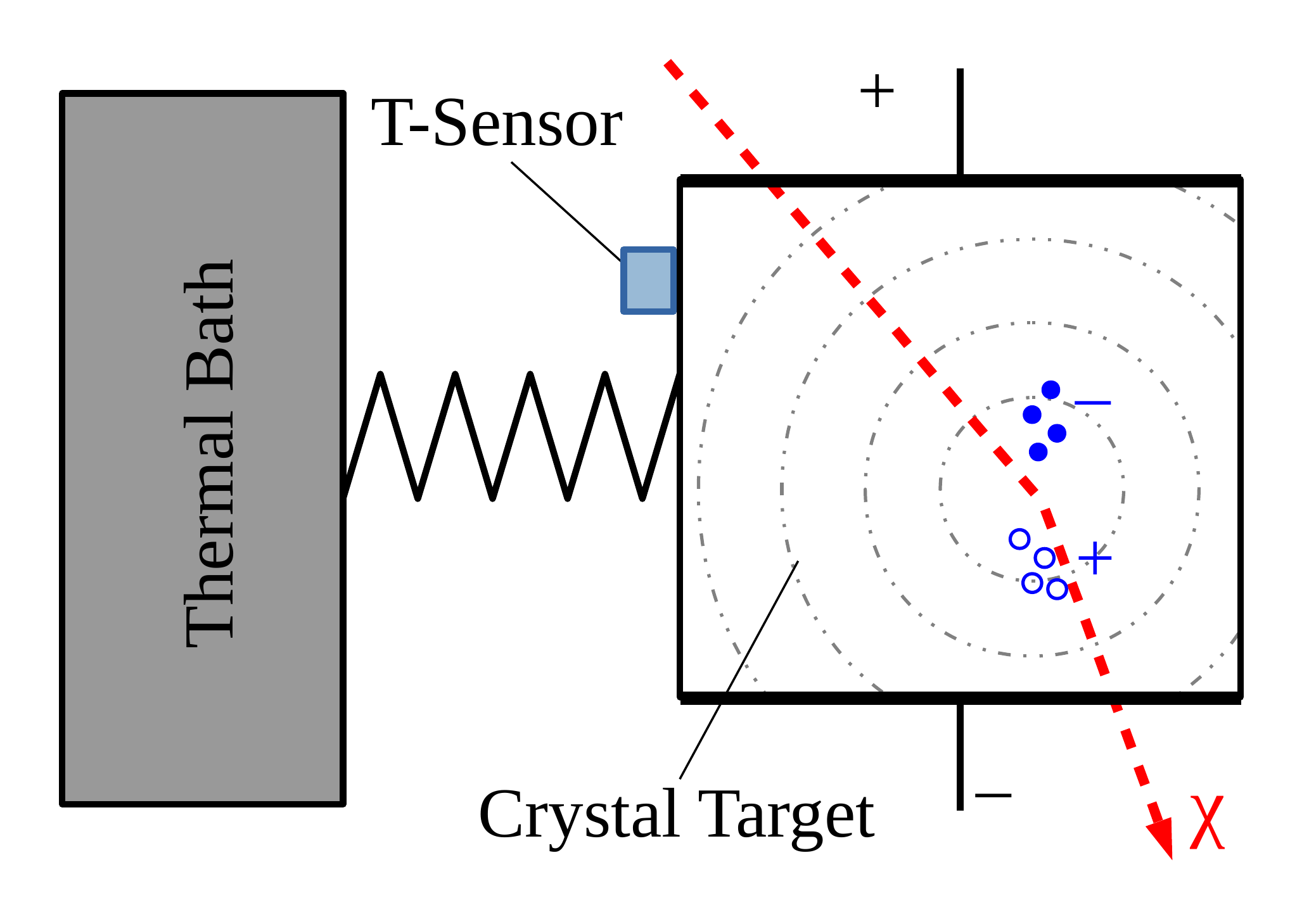} 
\caption{{\bf (Left)} The detector principle of scintillating crystals searching for an annually modulating signals is rather simple. However, extremely low intrinsic background levels are required. {\bf (right)} Cryogenic detectors are cooled down to mK-temperatures and weakly coupled to a thermal bath. Measured observables are heat (in form of phonons) and ionization (or scintillation light).}
\label{fig::crystals}
\end{figure}

In order to increase the target mass a different type of experiments uses arrays of high-purity scintillator crystals, mainly NaI(Tl) but also CsI(Tl). Their advantage is a rather simple detector design which can be operated stably for long periods of time: an array of kg-scale crystals making up a large target mass, read by low-background photomultipliers (see Fig.~\ref{fig::crystals}, left). The high mass numbers of~I ($A$\,=\,127) and Cs ($A$\,=\,133) lead to a high sensitivity to spin-independent interactions. The shortcomings of this detector design are a comparatively high intrinsic background level (the long-standing world-record for NaI-crystals is $\sim$1\,count/kg/d/keV~\cite{Bernabei:2018yyw}) which is enhanced by the inability to reduce backgrounds effectively by fiducialization or NR/ER discrimination (see Section~\ref{sec::backgrounds}). Therefore, experiments based on this technology mainly search for a possible dark matter-induced annually modulating low-energy signal above a constant contribution from backgrounds and the much larger non-modulating WIMP signal fraction. An event-by-event detection of dark matter candidates is not possible. Typical thresholds are 2\,keV$_\tnl{ee}$ (corresponding to 8\,keV$_\tnl{nr}$, 12\,keV$_\tnl{nr}$ and 22\,keV$_\tnl{nr}$ for Na, Cs and I, respectively) while results with  1\,keV$_\tnl{ee}$ have been published by DAMA/LIBRA recently~\cite{Bernabei:2018yyw}.  

\paragraph{B. Cryogenic Detectors} Crystalline detectors allow for the detection of the heat signal in form of phonons by measuring the temperature increase following a particle interaction. The sensitivity $\sigma$ to record the temperature increase from an energy deposition $E$ is given by 
\begin{equation}
\sigma^2 = c_1 \ kT \times (T \ C + c_2 \ E) \tn{,}
\end{equation}
where $T$ is the operation temperature, $C$ the detector's ($T$-dependent) heat capacity and $c_{1,2}\approx {\cal O}(1)$ are detector-specific constants describing the thermal coupling of the detector to the heat bath, readout noise, etc. To optimize the sensitivity, i.e., to reduce $\sigma^2$ it is thus mandatory to operate the detector at cryogenic temperatures, typically $\le$50\,mK, and to reduce the heat capacity $C$. Dielectric crystals, such as Ge or Si, are particularly well-suited for cryogenic operation as their heat capacity is given by $C \propto M \times T^3$ below their Debye temperature which is well above room temperature for Ge and Si. $M$~denotes the mass of the detector. One possibility to detect the tiny temperature rise $\Delta T$ are transition edge sensors (TES): they consist of thin wires (e.g., made from~W) which are operated at the transition temperature between their super-conducting and the normal-conducting state. A small temperature increase will have a big impact on the resistivity of the wire and thus the current running through it. An alternative detector-type are neutron transmutation doped (NTD) germanium thermistors whose resistivity strongly depends on the temperature~\cite{haller:1994}.

It was demonstrated already in the~90s that the simultaneous measurement of the ionization signal allows for signal/background discrimination in the WIMP search as the partition of the signal into the two channels depends on the recoil type~\cite{ref::shutt1992}. All modern cryogenic experiments follow this two-channel approach, where the second channel can be ionization or scintillation, see Fig.~\ref{fig::crystals} (right).

The main advantages of cryogenic detectors are their precise energy measurement with almost no quenching in the heat channel, their excellent energy resolution and the possibility to exploit two detection channels for excellent background rejection. However, the rejection breaks down at low energies around 5\,keV$_\tnl{nr}$, where the distributions start to overlap. The operation at mK-temperatures, typically using dilution refrigerators, is challenging and expensive and the mass of a single detector is limited to the kg-scale due to the requirement of a small heat capacity. This is overcome by using arrays of quasi-identical crystals, however, the surface-to-volume ratio of such experiment is not optimal and surface contamination has to be rejected. This is achieved by, e.g., optimizing the shape of the electric field across the crystals to distribute the ionization signal of surface events on the segmented readout electrodes differently to the signals from events happening in the bulk. This allows for the definition of an inner fiducial target~\cite{Broniatowski:2008yyl}. Rejection of surface events based on the rise-time of the phonon signal is employed as well: acoustic phonon pulses from events close to the detector surface rise faster than pulses from the bulk since impurities, defects etc.~on the surface enhance the conversion of the initially created optical phonons to the detected low-frequency acoustic phonons~\cite{Akerib:2007zz}.

In the recent years, the sensitivity of cryogenic HPGe detectors to low-mass WIMP ($\lesssim5$\,GeV/$c^2$) has been extended by operating the crystals at a relatively high bias voltage $V_\tnl{bias}$. This leads to a conversion of the charge signal into heat by the production of Neganov-Luke phonons~\cite{ref::luke1990} during the transport of the $n_e$ ionization electrons (and the corresponding holes). These phonons are detected on top of the primary heat signal $E_\tnl{heat}$ from the primary radiation,
\begin{equation}
 E_\tnl{dep} = E_\tnl{heat} + n_\tnl{e} \ e \ V_\tnl{bias} \tn{,}
\end{equation}
what allows reaching very low thresholds down to $\sim$50\,eV$_\tnl{ee}$. 
Since the initial ionization signal is not detected the ionization yield cannot be used for background rejection. However, the pulse shape carries some information on the position of the interaction in the detector.

\paragraph{C. Noble Liquid Detectors} The noble gases argon and xenon are excellent scintillators and can be ionized easily. Krypton has similar properties but is not used for dark matter searches because of its high intrinsic background from long-lived isotopes. Neon has a low mass number and was so for only used in gaseous state to search for low-mass WIMPs~\cite{Arnaud:2017bjh}. Both, argon and xenon can be liquefied to build dense and compact dark matter targets, with boiling points of 87.2\,K ($-$186.0$^\circ$C) and 165.2\,K ($-$108.0$^\circ$C), respectively. Both boiling points lie conveniently above the temperature of liquid nitrogen.

A particle interaction in the liquid noble gas target produces heat (which is not detected) as well as excited X$^*$ and ionized atoms X$^+$. The X$^*$ combines with neutral atoms~X forming excimer states X$^*_2$, which subsequently decay under the emission of ultraviolet light:
\begin{equation}\label{eq::scint}
\textnormal{X}^*  \xrightarrow{+\textnormal{X}} \textnormal{X}^*_2 \rightarrow 2\, \textnormal{X} + h \nu.
\end{equation}
The photons have a wavelength of 128\,nm and 178\,nm for argon and xenon, respectively. While sensitive photocathodes exist for the xenon scintillation light, this is not the case for argon. Before being detected by photosensors, the light thus needs to be shifted to optical wavelengths using a wavelength shifter such as tetraphenyl butadiene (TPB). This fluorescent material emits blue light at 425\,nm. 

\begin{figure}[b!]
\centering
\includegraphics*[width=0.495\textwidth]{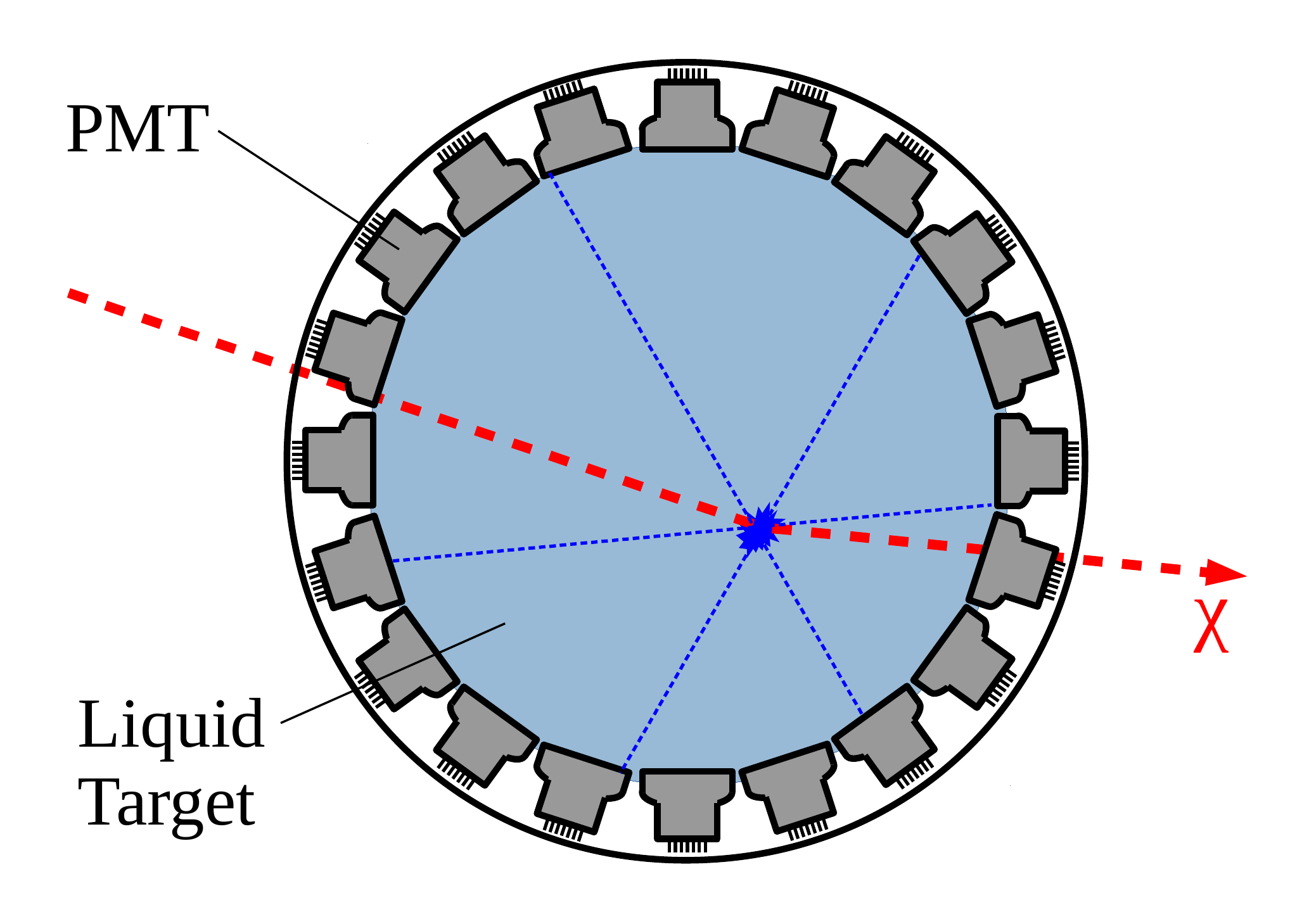}   
\includegraphics*[width=0.495\textwidth]{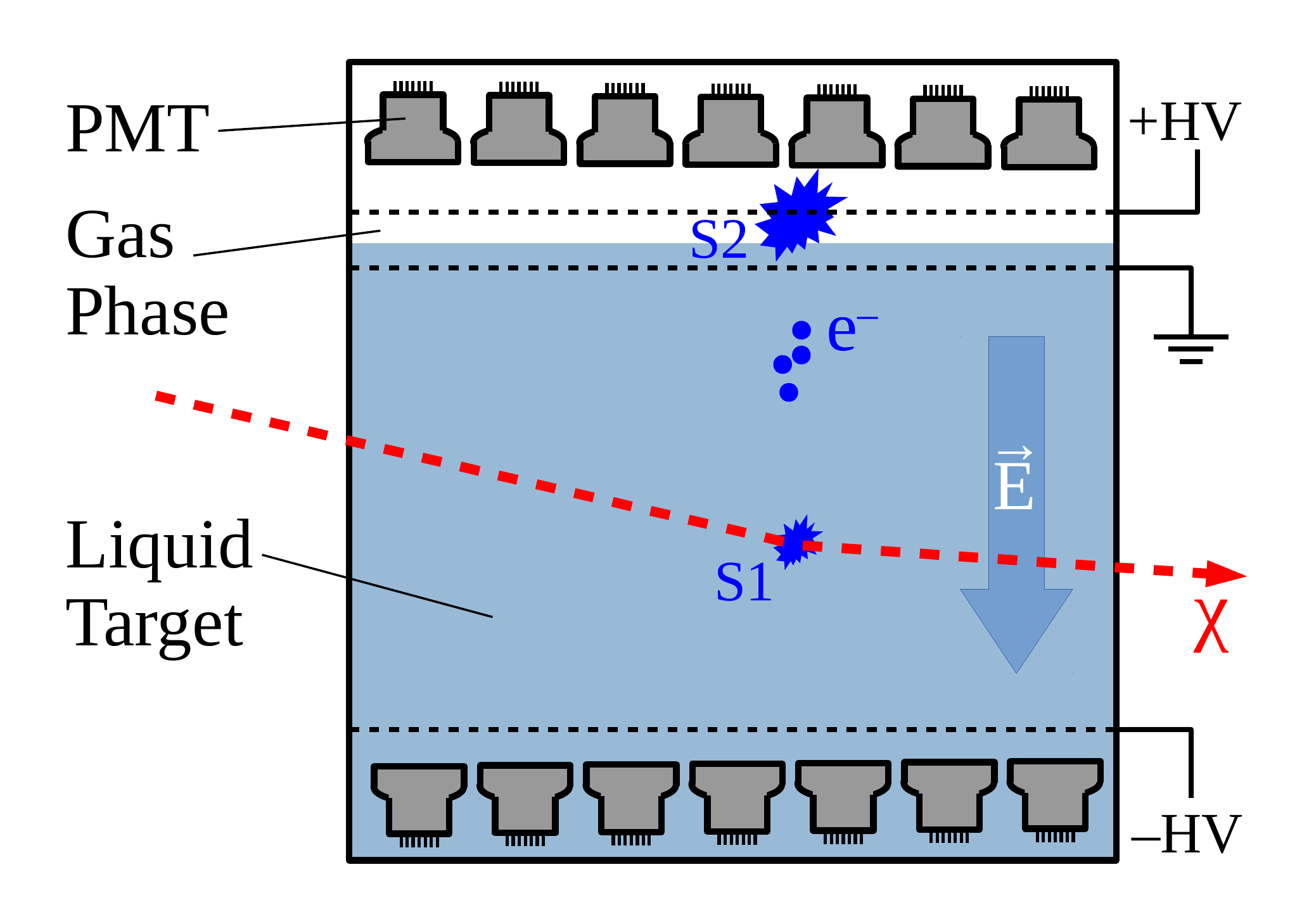} 
\caption{Detectors using the liquid noble gases argon or xenon as WIMP target either {\bf (left)} measure only the primary scintillation signal (single phase detectors) or {\bf (right)} detect the primary scintillation light as well as the ionization signal in a dual-phase time projection chamber (TPC). To achieve the lowest possible threshold, the ionization-induced charges are usually converted into a light signal via proportional scintillation after extracting the electrons into the gas phase above the liquid target.}
\label{fig::noble}
\end{figure}

Depending on the relative orientation of the two atomic spins in the excimer $X^*_2$, the scintillation light is emitted with different decay constants. With 4\,ns and 22\,ns, the singlet and triplet state lifetimes are very similar for xenon, however, they differ by almost three orders of magnitude for argon (7\,ns and 1.6\,$\mu$s). Since the relative population of singlet or triplet states depends on the ionization density and hence the interaction type, this leads to very different scintillation pulse shapes in argon and allows for the very efficient rejection of electronic recoil background to the $10^{-8}$~level (see Sect.~\ref{sec::backgrounds}). 
As natural argon gas contains the radioactive $\beta$-decaying isotope $^{39}$Ar ($T_{1/2}=269$\,y) at a level of $\sim$1\,Bq/kg, the high rejection is mandatory for argon-based WIMP searches. The use of $^{39}$Ar-depleted argon from underground sources reduces this problem by more than three orders of magnitude~\cite{ref::depleted}, however, it significantly increases the price of the gas. Since the Earth's atmosphere contains only a fraction of $9\times 10^{-8}$ of xenon it is also rather costly, however, it does not contain any long-lived radioactive isotope besides the double-beta decaying $^{136}$Xe ($T_{1/2}=2.12 \times 10^{21}$\,y).

The ions X$^+$ form singly ionized molecules X$^+_2$ with neutral atoms. In a time projection chamber (TPC) a charge measurement is possible by removing the ionization electrons from the interaction site by means of a strong electric field $\vec{E}_d$. Otherwise the electrons will recombine forming excited xenon atoms, which eventually will again decay via the emission of scintillation light:  
\begin{equation}\label{eq::ion}
\textnormal{X}^+ + e^- \xrightarrow{+\textnormal{2X}} \textnormal{X}^+_2 + \textnormal{X} \xrightarrow{+e^-} 2\,\textnormal{X} + \textnormal{X}^{**} \to 2\,\textnormal{X} + \textnormal{X}^* + \textnormal{heat} \xrightarrow{\textnormal{(\ref{eq::scint})}} 4\, \textnormal{X} + h \nu.
\end{equation}
The recombination fraction depends on the strength of the electric field $\vec{E}_d$. Scintillation and the ionization signal are anti-correlated: a simultaneous measurement improves the energy resolution of a detector~\cite{ref::ces}.

Two types of detectors filled with a liquid noble gas target are employed and ton-scale experiments of both types are being operated: \emph{single-phase} detectors measure solely the scintillation signal. Their spherical noble-liquid target is surrounded by photomultipliers in a $4\pi$ geometry to achieve a very high light yield, see Fig.~\ref{fig::noble} (left). While single-phase argon detectors employ pulse-shape discrimination background reduction in xenon detectors is limited to target fiducialization, i.e., defining a clean inner part of the active volume. The interaction position is defined based on photon timing and the signal distribution in the photosensors with a typical position resolution of a few~cm. 

\emph{Dual-phase TPCs} also record the primary scintillation signal (usually called S1) by means of two  arrays of photomultiplier tubes (PMTs) installed above and below a cylindrical noble-liquid dark matter target, see Fig~\ref{fig::noble} (right). The upper PMT array is placed in the gaseous phase above the liquid. The electric drift field ($\vec{E}_d =$\,0.1-1.0\,kV/cm) created across the liquid target by wire or mesh electrodes removes the ionization electrons from the interaction site and drifts them towards the gas phase on top of the detector. A second field (``extraction field'', $\sim$10\,kV/cm) across the liquid-gas interface pulls the electrons into the gas phase where they produce a secondary scintillation signal (S2) by collisions with the gas atoms. The S2~signal is proportional to the number of electrons and measured with the PMTs as well. The simultaneous measurement of both signals allows for identification of multiple scatters and the reconstruction of the location of the interaction with mm-precision. Thanks to the high target densities ($\rho_\tnl{Xe}=2.94$\,g/cm$^3$, $\rho_\tnl{Ar}=1.40$\,g/cm$^3$ at their boiling points) backgrounds are efficiently reduced by target fiducialization. The partition into excitation and ionization depends on the ionization density and the ratio S2/S1 can thus be used to distinguish electronic from nuclear recoils. Compared to other methods, the rejection levels are only fair reaching $\sim$1\,$\times$\,10$^{-3}$ at 30\% nuclear recoil acceptance. Liquid argon-based detectors are thus mainly relying on background rejection based on the pulse-shape of the scintillation signal~S1.

The detector threshold of dual-phase TPCs is determined by the scintillation light signal and can be greatly reduced in a charge-only (``S2-only'') dark matter search~\cite{Angle:2011th}. Here, no detection of an S1~signal is required and threshold around 1\,keV$_\tnl{nr}$ can be achieved. However, backgrounds are significantly higher since one loses the possibility to reject electronic recoils and the $z$-position of the event which reduces the power of fiducialization. The same data can also be analyzed in terms of sub-GeV dark matter scattering off electrons~\cite{Essig:2012yx}, see Sect.~\ref{sec::er}.

\paragraph{D. Bubble Chambers} Superheated liquids, usually refrigerants such as CF$_3$I, C$_3$F$_8$, C$_4$F$_{10}$, C$_2$ClF$_5$ or C$_3$ClF$_8$, are used as WIMP target in bubble chambers. The liquids are kept at a temperature just below their boiling point. A sufficient energy deposition into a certain micro-volume will lead to a local phase transition of the superheated liquid and start the formation of a bubble. The probability for bubble formation depends on the specific energy loss d$E$/d$x$ of the recoiling particle and can be tuned such that only nuclear recoil events from $\alpha$-particles, neutrons or WIMPs can create bubbles. The detector is then almost immune to the usually dominant electronic  recoil background sources ($\gamma$, $\beta$) which can be rejected by a factor $<$10$^{-9}$. $\alpha$-particles can be rejected based on the detailed acoustic characteristics of the bubble ``explosion''~\cite{ref::acoustic}.

\begin{figure}[t!]
\centering
\includegraphics*[width=0.495\textwidth]{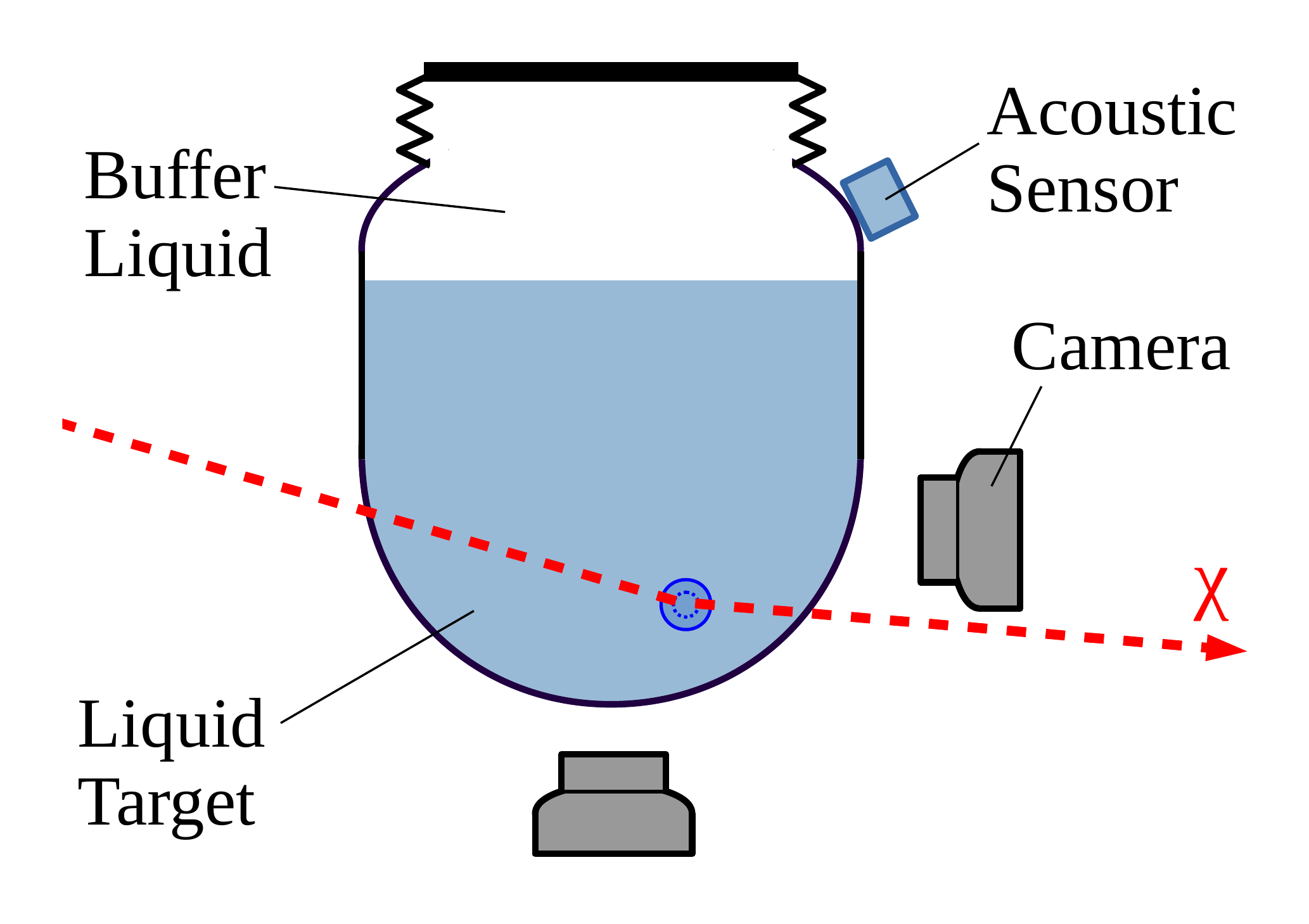}   
\includegraphics*[width=0.495\textwidth]{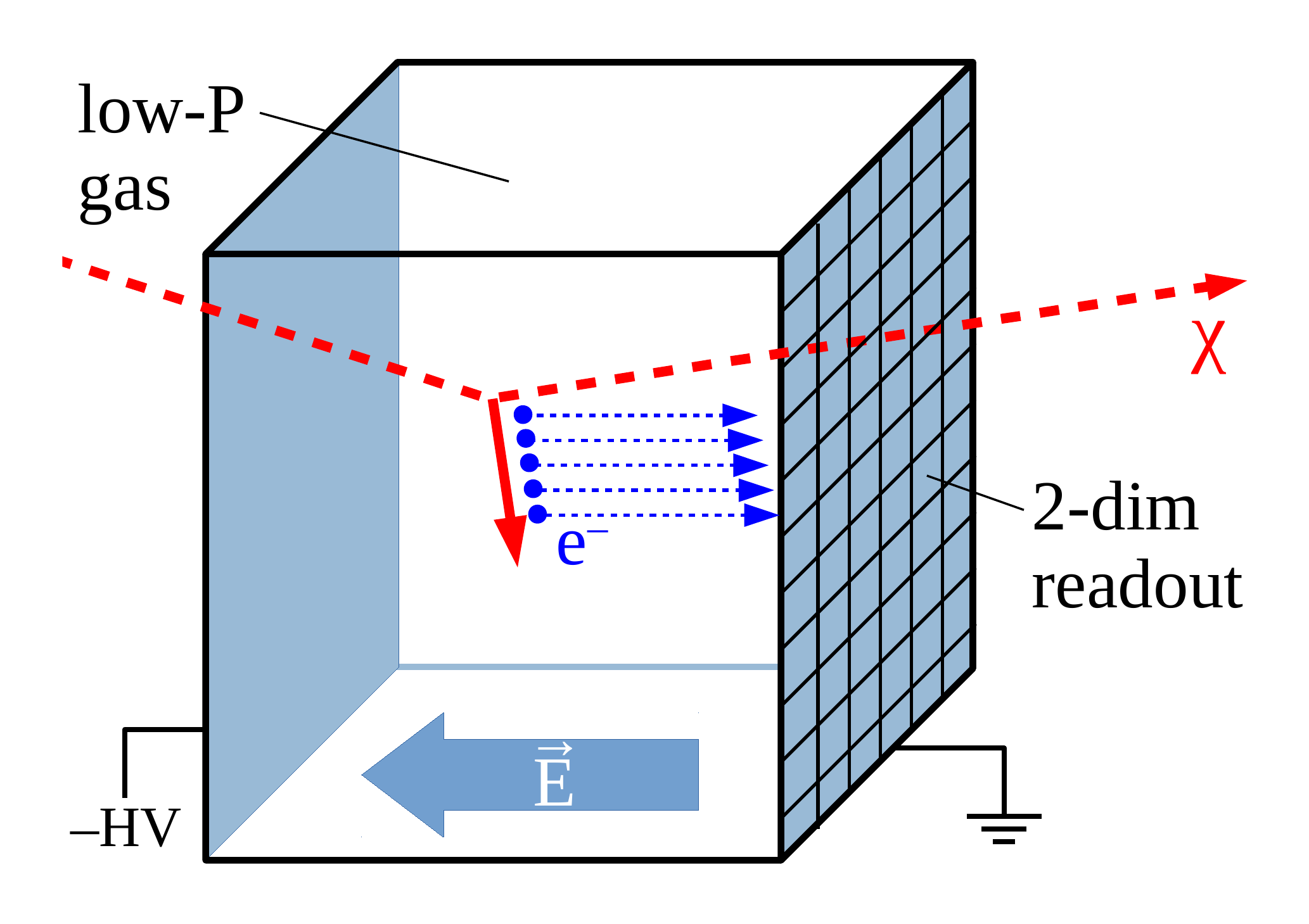} 
\caption{{\bf (Left)} The expected signal from a WIMP interaction in a bubble chamber is a single bubble in the superheated fluid which is recorded photographically. {\bf (Right)} Directional detectors, such as a TPC filled with a low-pressure gas and a high-granularity readout at the anode as illustrated here, aim at detecting the direction of the nuclear recoil signal.}
\label{fig::bubble}
\end{figure}

After each event, the bubble chamber has to be compressed to remove the bubble and decompressed to bring the liquid back into the superheated state. This involves mechanical motion and induces a rather long detector deadtime and complicated detector calibration. As in the early ages of particle physics, the (typically cylindrical) bubble chambers are read out by means of cameras, which are usually arranged in a stereoscopic way around the target, see Fig.~\ref{fig::bubble}~(left). The image allows the 3d-reconstruction of the event position with mm-resolution as well as measuring the multiplicity of the event.

The direct reconstruction of the recoil energy is not possible in bubble chambers as every energy deposition above the nucleation threshold will lead to a bubble. They are thus operated as threshold detectors measuring the integral rate above a certain minimal energy, which can be adjusted by controlling the pressure and temperature of the used fluid. This is calculated using the Seitz-model~\cite{ref::seitz}; thresholds down to 3.3\,keV were already achieved~\cite{ref::pico2016}. Superheated droplet detectors rely on the same detection principle. Their deadtime is reduced compared to bubble chambers as the bubbles are trapped in a water-based polymer matrix~\cite{ref::picasso}; however, bubble chambers were realized with much larger target masses at the 50\,kg-scale~\cite{Amole:2017dex}. 

The possibility to operate the bubble chamber with various target fluids of different composition makes them very versatile instruments for the WIMP search. Almost all used target fluids contain the isotope $^{19}$F, which has the highest sensitivity to spin-dependent WIMP-proton couplings (see Tab.~\ref{tab::sd} in Sect.~\ref{sec::rates}) and makes bubble chambers the leading technology for this channel. The possibility to add the heavy element iodine ($A\approx127$) offers a good sensitivity to spin-independent interactions. 

\paragraph{E. Directional Detectors} The Sun is moving around the Galactic center such that the average direction of the ``WIMP wind'' through the solar system originates from the constellation of Cygnus, see Section~\ref{sec::rates}. If the direction (or at least an head-tail asymmetry) of nuclear recoil events would be detected, a statistical discrimination of a WIMP signal from the uniformly distributed background would be possible if $\sim$30\,WIMP events are observed~\cite{Copi:1999pw}. Since the Earth rotates once a day with respect to the expected direction of the WIMP flux, the detectors either observe a daily modulation of the WIMP direction, or they need to be fixed on a parallactic mount to compensate for the rotation. If detectors have track reconstruction capabilities but integrate over long periods of time without following the Earth's rotation, the preferred WIMP direction is preserved, however, more than double the number of signal events is required to detect dark matter~\cite{OHare:2017rag}. 

As the nuclear recoil's track length depends on the target density, and a longer track facilitates the reconstruction of the track direction, most directional detectors feature low-pressure gas targets ($\sim$40-100\,mbar) with either photographic or a fine-granularity track readout in a TPC geometry, see Fig.~\ref{fig::bubble}~(right). The large number of channels required to reconstruct the short tracks makes the detector readout rather costly. The most common target gas used is CF$_4$, which provides sensitivity to spin-dependent WIMP-nucleon interactions; sometimes it is used in mixtures with other gases (e.g., CS$_2$, CHF$_3$). The typical threshold of existing detectors is around 20\,keV$\tnl{ee}$, while a competitive low threshold of 2\,keV$\tnl{ee}$ has been achieved in the MIMAC prototype~\cite{ref::mimac}. MIMAC also demonstrated an angular resolution of $14^\circ$~for 6.3\,keV$\tnl{nr}$~recoils~\cite{Tao:2019wfh}; the typical numbers achieved by other detectors are considerably higher.

The most sensitive directional detector is the 1\,m$^3$-scale DRIFT-II TPC~\cite{Battat:2016xxe}. Its target comprises 0.140\,kg when filled with 55\,mbar of a CS$_2$+CF$_4$+O$_2$ mixture. Due to the electronegativity of the CS$_2$ gas, the electrons generated by a particle interaction create negative ions which are drifted towards a multi-wire proportional chamber (MWPC) readout plane. This  significantly reduces diffusion and allows for fiducialization of the target. Electronic recoil background can be rejected to high levels ($2 \times 10^{-7}$) based on their longer range and lower ionization density. The achieved limit of $2.8 \times 10^{-37}$\,cm$^2$ at $m_\chi=100$\,GeV/$c^2$ for spin-dependent WIMP-proton couplings from a 7.66\,kg\,$\times$\,d run of DRIFT-II has to be compared with the current best limit in this channel from PICO-60, a bubble-chamber filled with 52\,kg of C$_3$F$_8$. An efficiency corrected exposure of 2571\,kg\,$\times$\,d led to a 3.5\,orders of magnitude more sensitive upper limit at the same mass~\cite{Amole:2019fdf}. Ignoring thresholds, backgrounds and the different fraction of $^{19}$F in the two targets, and assuming that the WIMP would have a cross-section at the current best upper limit, the directional detector would need to be $\sim$10$^4$\,times larger (or measure $\sim$10$^4$\,times longer) to detect an excess of $\sim$30 signal events. The prospects for directional detectors are more promising at low WIMP masses $m_\chi\lesssim5$\,GeV/$c^2$, where the neutrino floor (see Sect.~\ref{sec::floor}) becomes relevant at cross sections four orders of magnitude higher than for larger $m_\chi$: measuring the direction of a nuclear recoil provides the possibility to distinguish the neutrino background from the Sun from a WIMP signal~\cite{Grothaus:2014hja}.

\subsection{Backgrounds and Background Mitigation Strategies}
\label{sec::backgrounds}

The search for WIMP dark matter, with its featureless exponentially falling recoil spectrum as expected for spin-(in)dependent WIMP-nucleus scattering\footnote{Other interaction models predict different signatures, but no model expects a ``smoking gun''-signature like a monoenergetic line.}, requires an extremely low background in the energy region of interest (ROI). Usually, the design of the experiments aims for ``zero background'' ($\lesssim$1\,events in the ROI for the design exposure) such that the observation of a few signal events would already have a high statistical significance. In this section we will review the most common background sources and background mitigation strategies.

\subsubsection{Background Sources}

There are various possibilities to classify the sources of background in dark matter experiments. We chose to organize them around the signature they leave in the detector: 

\paragraph{Electronic Recoil Background} 
The overall background of most current WIMP search experiments is dominated by $\gamma$-back\-grounds from the environment or the detector setup itself and by $\beta$-particles at the surfaces or in the bulk of the detector. The particles generate electronic recoils (ER) by electromagnetic interactions with the atomic electrons. The backgrounds originate from long-lived natural radioisotopes such as the elements of the primordial $^{238}$U and $^{232}$Th chains and $^{40}$K, but also from anthropogenic isotopes in the environment such as $^{85}$Kr, $^{137}$Cs or $^{110m}$Ag which stem from nuclear fuel element recycling plants and accidents. $^{60}$Co is usually present in steel components; grade-1 titanium has often a lower radioactivity level at only little weaker material strength~\cite{Akerib:2017iwt}. Oxygen-free high-conductivity (OFHC) copper is considered the best construction material in terms of background, however, it is rather soft and has a limited mechanical strength. 

Another source of $\beta$- and $\gamma$-backgrounds are isotopes in the material of the shielding or the detector itself which were produced by cosmic ray activation in proton or neutron-capture reactions. A prominent example with a very long half-life of $T_{1/2}=269$\,y is $^{39}$Ar which is produced in the atmosphere by cosmic ray neutrons in the reaction $^{40}$Ar($n$,$2n$)$^{39}$Ar or underground by muon ($^{39}$K($\mu^-$,$\nu_\mu$)$^{39}$Ar) or neutron capture ($^{39}$K($n$,$p$)$^{39}$Ar) on $^{39}$K~\cite{Sramek:2015bfa}. In other cases, the resulting isotopes are short-lived ($T_{1/2}<1$\,y) and underground-storage of the materials prior to the dark matter search reduces their impact on the experiment. Nevertheless, to build a dark matter experiment it is important to procure freshly produced materials (e.g., Cu, Ar, Xe, Ge-crystals) with a low activation whenever possible and to optimize the transportation of the materials to the laboratory. Air transport is often avoided as exposure at high elevations is significantly worse than at sea level~\cite{Baudis:2015kqa}. At the current scale and background level, no experiment has to be worried about underground activation induced by cosmogenic muons (which themselves are never a problem since they depot huge signals), however, this potential background needs to be studied for the ultimate detectors (see Section~\ref{sec::floor}).

A special background isotope is $^{222}$Rn\footnote{The argumentation is similar for $^{220}$Rn from the $^{232}$Th chain, however, its background contribution is lower thanks to its shorter half-life and reduced abundance in detectors.}, a member of the $^{238}$U chain with the long-lived mother $^{226}$Ra. As a noble gas, it is constantly emanated from all surfaces and thus present in most experiments, especially if they use gas or liquid targets. Its $\alpha$-decay deposits a huge energy and is usually not contributing to the dark matter background directly, however, $\beta$-decays of its daughter $^{214}$Pb to the ground state of $^{214}$Bi, which are not accompanied by a prompt $\gamma$-emission, can lead to single scatter ER events in the WIMP search energy window. In addition, radon plate-out on surfaces, i.e., the collection of $\alpha$-decaying daughters, can lead to background if most of the $\alpha$-energy of the subsequent decays are lost in the surface. With a specific activity of only 0.18\,$\mu$Bq per kg of target material the lowest $^{222}$Rn concentration in a dark matter experiment so far was achieved by DEAP-3600~\cite{Amaudruz:2017ekt}

Even though ER backgrounds are usually dominating the overall background budget, all high-sensitivity, low-background detectors have means to reject ER signals at a very high efficiency (see discussion below in Sect.~\ref{sec::mitigation}). This usually renders ER backgrounds sub-dominant for the actual WIMP search. The ultimate ER background will be elastic collisions of low-energy solar pp- and $^7$Be neutrinos with atomic electrons~\cite{Baudis:2013qla} as this background source can neither be avoided nor shielded but only rejected in the analysis based on their ionization density.  

\paragraph{$\alpha$-Background} 
$\alpha$-contamination in the dark matter target or the detector materials facing the target is usually uncritical due to the large energy deposited in the detector. It only becomes relevant if the major part of the $\alpha$-energy is lost in insensitive detector regions (e.g., detector wall), producing artifacts which can leak into the signal region (for an example see~\cite{Strauss:2014aqw}). A notable exception are bubble chambers and some directional detector concepts as described in Section~\ref{sec::detectors}~D, where $\alpha$-particles are the main source of background since the detectors are insensitive to ER signals.

\paragraph{Nuclear Recoil Background}
The most critical background for direct detection experiments are single-scatter neutron-induced nuclear recoils (NRs) from ($\alpha,n$) and spontaneous fission reactions (radiogenic background), or NRs induced by cosmic ray muons (cosmogenic background), as such events cannot be distinguished from a WIMP signal. For this reason, NRs induced by the coherent scattering of neutrinos off the target nuclei will eventually constitute the ultimate background for direct WIMP searches (see Sect.~\ref{sec::floor}) as there is no possibility to reject this background on an event-by-event basis. Compared to $\gamma$-rays of the same energy, neutrons have a significantly longer mean free path (e.g., {\cal O}(10)\,cm in liquid xenon) and are thus much harder to shield. However, in larger detectors they have a rather high probability to scatter elastically off the target nuclei several times creating a non WIMP-like multi-scatter signature.

\emph{Radiogenic neutrons} can originate from \emph{$(\alpha,n)$ reactions} where an $\alpha$-particle emitted by a radioactive decay, usually from the primordial decay chains, is absorbed by a nucleus and knocks off a neutron. A well-known example for this process is the $^{241}$AmBe neutron source, which is employed for calibration purposes in many dark matter detectors. Due to the short path length of the $\alpha$-particle, the process is only relevant if the $\alpha$-emitter is mixed into a material with a high $(\alpha,n)$ cross section. The reaction probability depends on its $Q$-value, the difference of the $(\alpha,n)$ threshold energy and the $\alpha$-energy as well as on the Coulomb barrier which effectively suppresses the $(\alpha,n)$ process for heavy nuclei. Therefore, the neutron production via $(\alpha,n)$ reactions is highest for low-$Z$ materials, typical examples being polyethylene ((C$_2$H$_4)_n$) or PTFE (Teflon, (C$_2$F$_4)_n$). The latter is often used in dark matter detectors and contains $^{17}$F with a particularly high $(\alpha,n)$ cross section. Another isotope to be avoided due to its high cross section is $^9$Be which typically occurs in springs made from CuBe bronze etc.

Since their high Coulomb-barrier prevents $(\alpha,n)$ reactions, very heavy nuclei ($>^{232}$Th) can essentially only generate neutrons via \emph{spontaneous fission}.  The highest contributions usually come from $^{238}$U and $^{235}$U present in materials. The spontaneous fission neutron yield normalized to the activity of the $^{238}$U decay chain is of the order of $6 \times 10^{-7}$\,$n$/decay; the neutrons have MeV-energies.

\emph{Cosmogenic neutrons} are created by fast muons with mean energies above 40\,GeV that make it into the underground laboratories (depth $>100$\,m.w.e.) where they interact with the detector and the surrounding environment. The produced neutrons have energies up to the GeV-range and thus have a great penetration power: even neutrons generated inside the rock a few meters away from an underground cavern can enter the experiment. The neutrons are produced in hadronic (had) or electromagnetic showers (em); in the latter case, the main process is photoneutron production by real photons in giant dipole resonances. A smaller fraction of the neutrons is produced by virtual photons (v). Muon-capture reactions are irrelevant at the underground depths considered here. The neutron yields $Y$ of the different processes are ordered in the following way~\cite{Malgin:2017dwh}
$$
Y_\tnl{had} > Y_\tnl{em} \gg Y_\tnl{v} \textnormal{.}
$$
The main production processes in the dominant hadronic showers are deep-inelastic $\pi N$ and $\pi^- N$ capture reactions.

\subsubsection{Background Mitigation}
\label{sec::mitigation}

In order to be sensitive to the rare signal from WIMP interactions inside a detector, the contribution from the various backgrounds has to be minimized. There are a plethora of different reduction strategies which are commonly dubbed ``low-background techniques'' (for a review see~\cite{ref::heusser}). All sensitive experiments make use of most of them in one or another way. 

\begin{figure}[b!]
\centering
\includegraphics[width=0.8\textwidth]{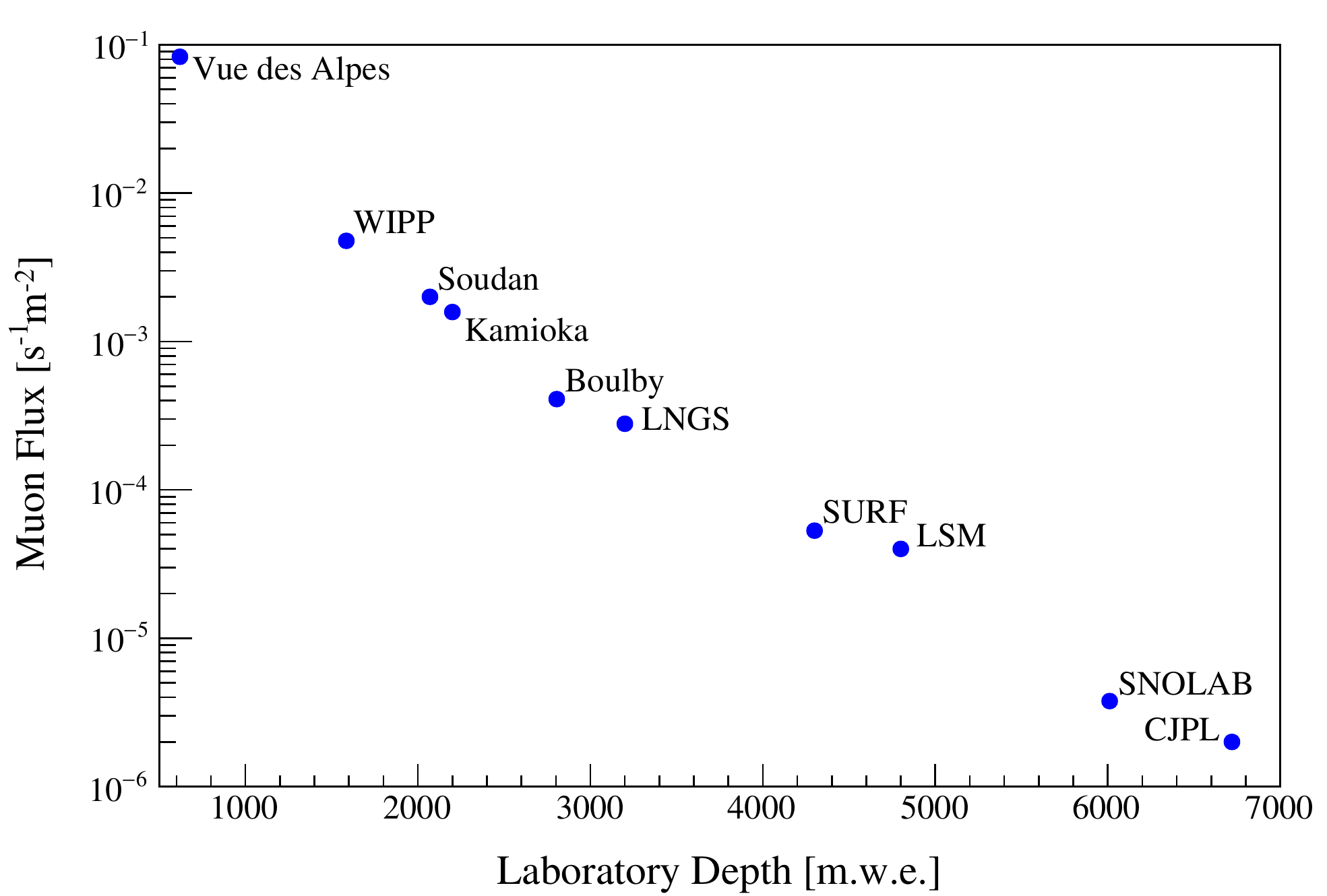}
\caption{Muon flux vs.~depth of various underground laboratories (measured in ``meters water equivalent'' (m.w.e.). While the depth increases from 620\,m.w.e.~to 6720\,m.w.e, the muon flux decreases by more than four orders of magnitude. } \label{fig::labs}
\end{figure}

\paragraph{Shielding}
Massive shields installed around the dark matter detector are used to suppress backgrounds originating from the experiment's surroundings. High-$Z$ materials such as lead and copper or large amounts of water are efficient against external $\gamma$-rays. Dense materials with a high hydrogen content (e.g., polyethylene or paraffin) or water efficiently moderate and eventually absorb neutrons. In order to reduce muon-induced neutrons, dark matter detectors are installed in deep-underground laboratories. To allow for the direct comparison between different laboratories, their effective depth is expressed in ``meters water equivalent'' (m.w.e.), see Fig.~\ref{fig::labs}. The hadronic cosmic ray component ($p$, $n$) is already blocked at depths $\gtrsim$10\,m.w.e. Above a few hundred m.w.e., the remaining muon intensity is reduced to below the level of radiogenic neutrons~\cite{ref::heusser}. At the typical rock overburden of $2\ldots6 \times 10^{3}$\,m.w.e.~of the laboratories hosting dark matter detectors, the muon flux is suppressed by 5-7\,orders of magnitude compared to the sea level flux. 

As many background sources predominantly lead to events close to the detector surfaces, their contribution can be further reduced by exploiting the self-shielding capability of the target material. This \emph{fiducialization} is especially effective for high-$Z$ detector materials and relies on the definition of an inner detector volume with reduced backgrounds. It is either based on the measurement of the interaction position of every single event with some precision, or by designing the detector in a way such that only events away from the walls will create a measurable signal. The latter approach is successfully employed by Ge-based experiments~\cite{Broniatowski:2008yyl}. In some cases, the timing or pulse shape of the detector signal contains information on its place of origin inside the detector~\cite{Akerib:2007zz,Aalseth:2010vx}.

Muon-induced neutrons are further reduced by active muon vetoes, typically made of large sheets of plastic scintillator or water-based Cherenkov detectors. These systems detect the minimum ionizing muon with a high efficiency which allows for the rejection of every coincident signal in the detector as this might stem from a muon-induced neutron. The veto systems usually also have some sensitivity to detect hadronic or electromagnetic showers; this is important for cases where the shower is created outside of the veto and no muon can be detected.

\begin{figure}[b!]
\includegraphics[width=\textwidth]{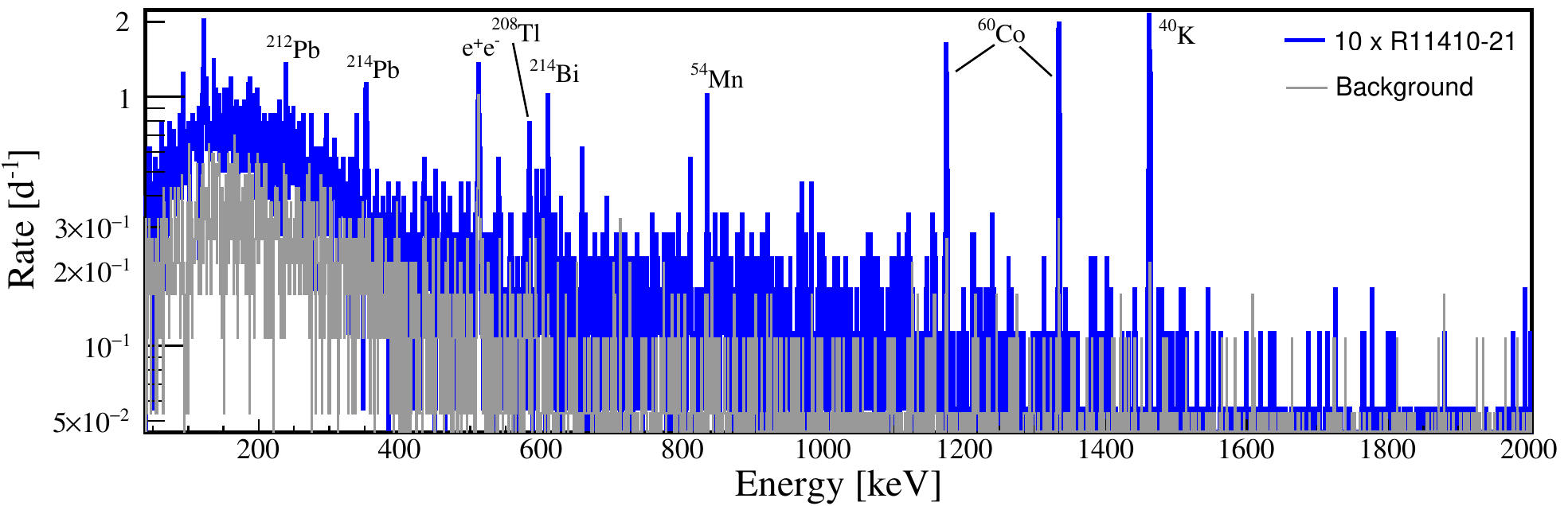}
\caption{$\gamma$-spectrum of a sample of 10~low-background PMTs (Hamamatsu R11410-21, used in LXe TPCs) to measure their intrinsic radioactive contamination, together with the background spectrum of the HPGe detector~\cite{Sivers:2016yvm}.  } \label{fig::hpge}
\end{figure}

\paragraph{Material Selection}
Thanks to the effective shielding, the ``external'' background from the detector surroundings (with the exception of cosmogenic neutrons) are irrelevant for almost every experiment. The main ER and NR backgrounds thus originate in the (active or passive) parts of the detector and the shield itself. They are reduced by using only materials and components with a low intrinsic radioactive contamination for detector construction. Several methods are available to identify such ``clean'' materials:
\begin{enumerate}
 \item $\gamma$-spectrometry: A non destructive method to measure the $\gamma$-activity of a material or detector component in a low-background HPGe detector (see Fig.~\ref{fig::hpge}). Instruments with large sample-cavities of {\cal O}(25)\,liters are available. The method is limited by the long measuring times required even for massive samples and the different sensitivity to different parts of the decay chains which complicates the detection of possible breaks of the radioactive equilibrium.
  \item Mass spectroscopy: Materials (mostly metals, but also some non-metals) are evaporated for glow discharge (GD-MS) or inductively coupled plasma mass spectrometry (ICP-MS). The methods can measure the concentration of elements to $\lesssim$ppt levels, however, the radioactive isotopes themselves remain undetected and one has to make assumptions on their natural abundance. In case of the U, Th-chains, the early part of the chain is measured (which is harder to assess with $\gamma$-spectrometry). 
  \item Neutron activation analysis: Small samples are irradiated by neutrons and the characteristic $\gamma$-lines of activated isotopes are measured in $\gamma$-spectrometers. The method gives access to isotopes which are not accessible by the other methods.
  \item Rn-emanation: The emanation of radon isotopes from materials is measured by collecting the daughters (e.g., $^{218}$Po, $^{214}$Po) on a PIN-diode where they undergo $\alpha$-decay. The sample is often placed in a large-volume chamber and the charged daughters are collected on the small diode via an electric field~\cite{ref::kiko2001}. Sensitivities of ${\cal O}$(100)$\mu$Bq/sample can be achieved. If the sample chamber is sufficiently large, the method is non-destructive.  
  \item $\alpha$-spectroscopy: The effect of Rn-plate out can be measured without destroying the sample with surface $\alpha$-counters.
\end{enumerate}

\paragraph{Target Purification}
In many cases the dark matter target has to be purified from radioactive impurities. In case of solid targets, this is done prior to the start of the experiment, e.g., during the process of growing crystals (e.g., NaI, CsI, CaWO$_4$, Al$_2$O$_3$, Ge). While pulling the crystal from the melt, the lattice will attract certain impurities and repel others. In the zone-refining procedure, part of a crystal is molten and slowly moved across the ingot, pushing the impurities to its end~\cite{ref::zonemelting}. This allows the production of crystals that are significantly cleaner than the pre-selected raw material. Cryogenic distillation exploits the different boiling points of target and background gases~\cite{ref::krcolumn} and is used to clean the noble gases argon and xenon from $^\tnl{nat}$Kr (which contains radioactive $^{85}$Kr at the $10^{-11}$-level). An alternative approach relies on gas chromatography in a charcoal column~\cite{ref::chromatography}. 

It is more complicated to remove the radioactive $^{39}$Ar which is present in an $^{\tnl{nat}}$Ar target at a concentration of 1\,Bq/kg as isotopic separation is required. The most efficient way to produce low-background argon gas is to extract it directly from underground wells where the $^{39}$Ar-level is reduced by more than three orders of magnitude~\cite{ref::depleted}. Gaseous or liquid targets can in principle be purified even after detector assembly, even on-line during data taking, however, for practical reasons and efficiency it is usually done before the start of an experiment. A notable exception is the active removal of Rn, which is constantly regenerated inside the detector by emanation from the walls. In this case, on-line removal is mandatory and very promising first results have been achieved by running a cryogenic distillation column for $^{85}$Kr removal from Xe in ``reverse mode''~\cite{Aprile:2017kop}. An alternative approach to reduce the background from Rn-emanation is the coating of surfaces to avoid that the isotope enters the active target.

\paragraph{Active Rejection}
While the methods discussed so far aim at avoiding backgrounds in the first place, there are always background events remaining that are recorded by the dark matter detector. These can be reduced by making use of the expected signature of a WIMP: a single-scatter, low-energy nuclear recoil\footnote{If one searches for non-standard WIMP-nucleon interactions predicted by models beyond spin-(in)dependent scattering, one, two or even all of these expectations might be subject to change.}. It is thus important to detect the event multiplicity (i.e., the number of interactions inside the detector) as WIMPs will only scatter once in the detector due to their tiny cross section. In particular neutrons with their longer mean-free path will often generate multiple-scatter signatures. Depending on the used technology, multi-scatter events can be identified either directly by measuring several interaction vertices or by pulse-shape studies. A less powerful but still viable alternative is to segment the detector in an array of smaller units and to reject events which are seen in several segments. All sensitive WIMP detectors can therefore identify and reject multi-scatter events.

\begin{figure}[t!]
\centering
\includegraphics*[width=0.495\textwidth]{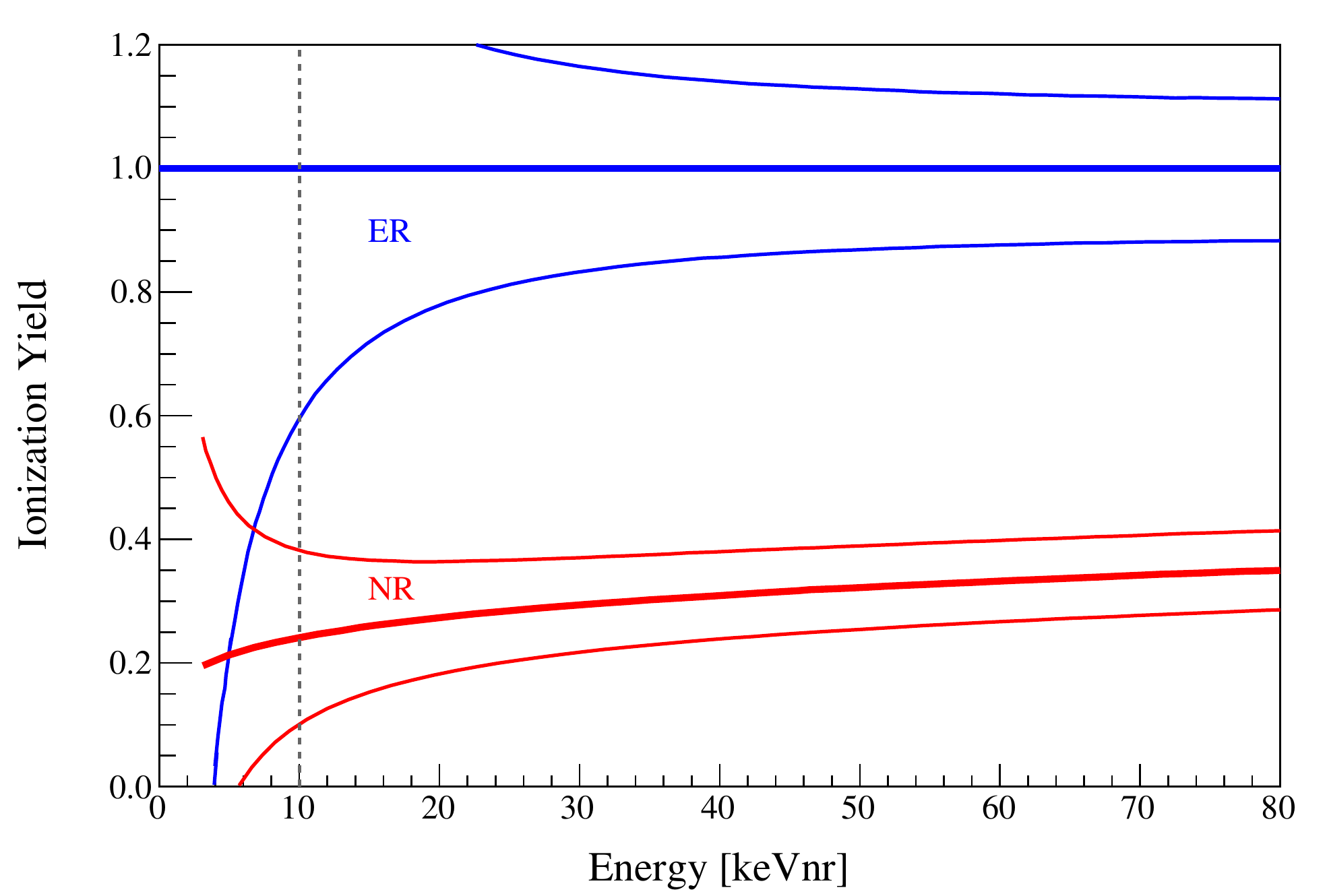}   
\includegraphics*[width=0.495\textwidth]{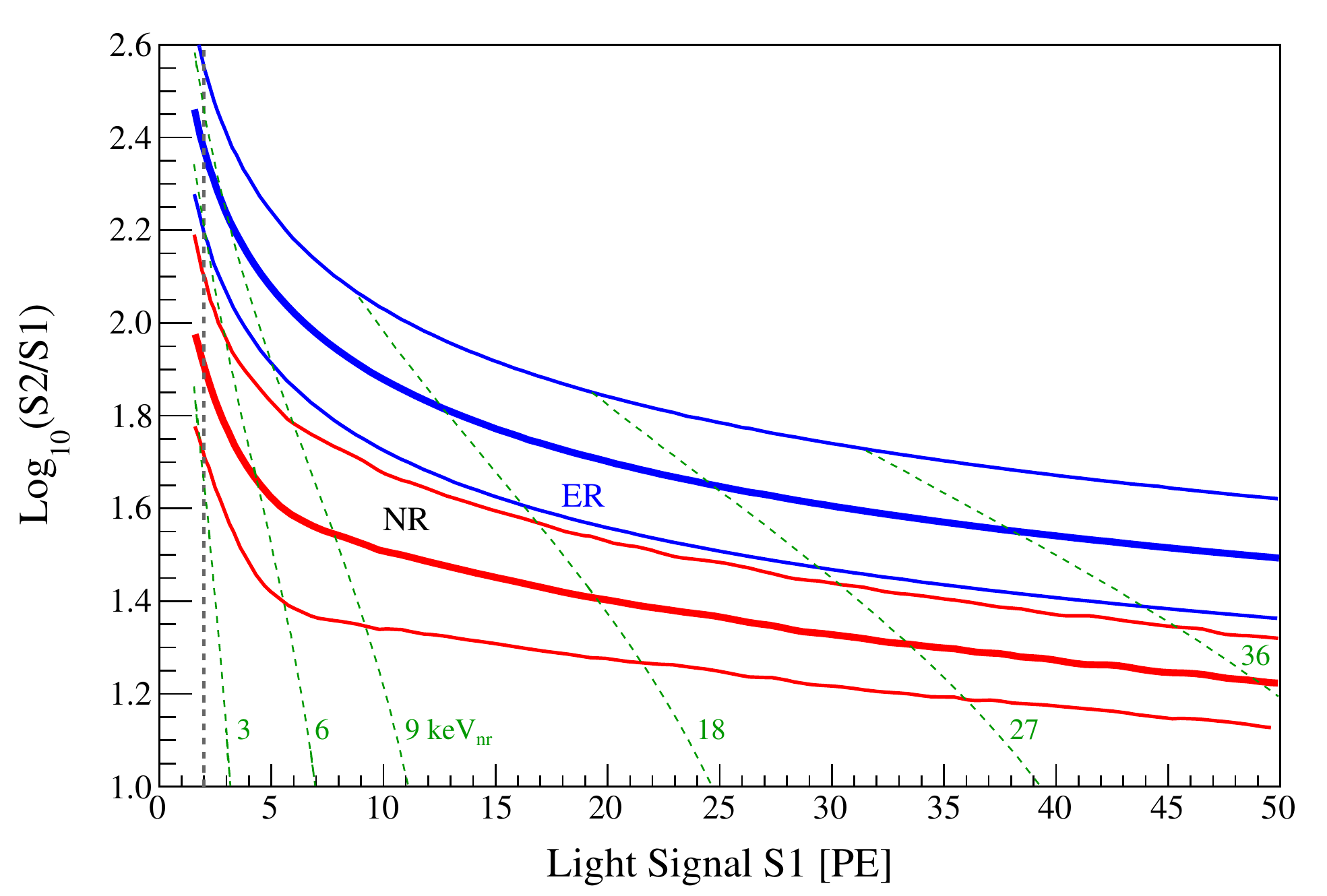} 
\caption{Examples for active rejection of electronic recoil backgrounds by measuring two independent signals in a detector. Both plots show the mean (thick) and the 10\% and 90\% quantiles of the distributions. The vertical dashed lines indicates the energy threshold used for the WIMP search. {\bf (Left)} Cryogenic experiments using germanium targets measure ionization and the phonon signal, which is used to set the energy scale (in keV$\tnl{nr}$). The ionization yield for nuclear recoil (NR) signals is reduced compared to electronic recoils (ER) and the distributions are well separated above $\sim$10\,keV$_\tnl{nr}$. Data from EDELWEISS-III~\cite{Armengaud:2017rzu}. {\bf (Right)} Liquid xenon time projection chambers measure scintillation light (S1) and ionization charge (S2) from an interaction. The ratio~S2/S1 is reduced for NR signals. The distributions are rather close together but ER rejection at reduced efficiency works down to the detector threshold. The energy scale is reconstructed from both signals (green lines). Data from LUX~\cite{Akerib:2013tjd}.}
\label{fig::rejection}
\end{figure}

The different ionization density of ERs compared to the WIMP-induced NRs is often used to reject ER signals. This effect has already been discussed in context of signal quenching in Sect.~\ref{sec::generation} and is exploited if two of the experimental observables charge, light and heat/phonons are available on an event-by-event basis as their ratio differs for ERs and NRs. For cryogenic experiments measuring phonons and another observable, this method very effectively rejects ERs ($<$10$^{-6}$ at a NR acceptance around 100\%~\cite{Armengaud:2017rzu,Agnese:2013ixa}), however, the ER and NR populations overlap at low energies, see Fig.~\ref{fig::rejection}. The method is less powerful for liquid noble gases, with ER rejection levels at the 99.9\% level at 30\% NR acceptance, but at least in xenon the ER discrimination works effectively down to lowest energies and does thus not impact the threshold (Fig.~\ref{fig::rejection}). In general, the achievable ER rejection factor is directly coupled to the NR acceptance: the first can be usually increased by lowering the latter. All experiments have to find ways to reduce the impact of events happening at or close to the detector surface where partial signal loss can lead to leakage of ERs into the NR region.

As mentioned above, the liquid noble gas excimer states $X^*_2$ de-excite with two different time constants, depending on whether the spins of the two atoms are in a singlet ($\uparrow \uparrow$) or a triplet configuration ($\uparrow \downarrow$). The relative population of the two states depends again on the ionization density $\rho_I$ and thus the recoil type, with less triplet states being produced for high $\rho_I$. This is effectively exploited in liquid argon, where the two time constants differ by three orders of magnitude: $\tau_{\uparrow \downarrow}=6$\,ns, $\tau_{\uparrow \uparrow}=1.5$\,$\mu$s~\cite{ref::hitachi}. By comparing the fast part of the scintillation signal $F_\tnl{prompt}$ to the total pulse area ERs can be rejected to a very high level ($<2 \times 10^{-8}$ at 90\% NR acceptance~\cite{ref::deap1}). However, the need for a large number of detected photons for \emph{pulse shape discrimination} increases the analysis threshold to $\sim$30\,keV$_\tnl{nr}$. For xenon, the two time constants are too similar to be exploited for ER rejection in a meaningful way~\cite{Hogenbirk:2018zwf}.

Bubble chambers operated at the right temperature regime are almost immune (at the $10^{-9}$~level) to interactions from particles producing ERs with a low ionization density as these do not create detectable bubbles. The remaining $\alpha$-background can be reduced by recording the acoustic signal from the bubble nucleation~\cite{ref::acoustic}: bubbles from $\alpha$-interactions sound different (louder) than the ones from NRs from neutrons (or WIMPs) due to the formation of multiple bubbles along the $\alpha$-track. Typical rejection levels are $\sim$99\% at $\sim$95\% signal acceptance.

\section{Direct Detection: Status and Future}
\label{sec::status}

Since the first attempt to experimentally detect cold WIMP dark matter using a standard 0.72\,kg high purity germanium crystal in 1987~\cite{Ahlen:1987mn}, numerous experiments utilizing the techniques discussed in Sect.~\ref{sec::detectors} have published their results. Apart from a few ``anomalies'' (see, e.g., \cite{Gresham:2013mua}), which usually faded away within a few months after triggering lots of theoretical activities to explain the result, no experiment observed a statistically significant excess above its background expectation. As of today, the WIMP still remains elusive. 

Here we focus on the current status of the WIMP direct detection landscape. The leading experiments are summarized in Table~\ref{tab::experiments}.

\begin{table}[t]
\caption{Alphabetical list of some of the leading direct detection experiments that published results on WIMP interactions; some of them are not operational anymore. The quoted mass refers to the active target; some detectors have position sensitivity and can select the (cleaner) innermost part of the detector for the analysis. Please note that the mass-range covers 7~orders of magnitude. The references list the latest or most relevant publication from the experiment; projects marked with an asterisk~($^*$) indicate prototype phases, i.e., the quoted mass does not represent the final stage. }\label{tab::experiments}
\centering
\small
\begin{tabular}{lllclr}
\hline \hline
Experiment & Type & Target & Mass [kg] & Laboratory & Ref. \\ 
\hline
ANAIS-112 & Crystal & NaI & 112 & Canfranc & \cite{Amare:2019jul} \\
CDEX-10  & Crystal & Ge & 10 & CJPL & \cite{Jiang:2018pic} \\
CDMSLite & Cryogenic & Ge & 1.4 & Soudan & \cite{Agnese:2015nto} \\
COSINE-100 & Crystal & NaI & 106 & YangYang & \cite{Adhikari:2018ljm} \\
CRESST-II & Cryogenic & CaWO$_4$    & 5 & LNGS & \cite{Angloher:2015ewa} \\
CRESST-III  & Cryogenic & CaWO$_4$ & 0.024 & LNGS & \cite{Abdelhameed:2019hmk} \\
DAMA/LIBRA-II & Crystal & NaI & 250 & LNGS & \cite{Bernabei:2018yyw} \\ 
DarkSide-50 & TPC & Ar & 46 & LNGS & \cite{Agnes:2018ves} \\
DEAP-3600 & Single phase & Ar & 3300 & SNOLAB & \cite{Ajaj:2019imk} \\
DRIFT-II  & Directional &  CF$_4$ & 0.14 & Boulby & \cite{Battat:2016xxe} \\ 
EDELWEISS & Cryogenic & Ge & 20 & LSM & \cite{Hehn:2016nll} \\
LUX     & TPC & Xe & 250 & SURF & \cite{Akerib:2016vxi} \\
NEWS-G  & Gas Counter & Ne & 0.283 & SNOLAB & \cite{Arnaud:2017bjh} \\
PandaX-II & TPC & Xe & 580 & CJPL & \cite{Cui:2017nnn} \\ 
PICASSO & Superheated Droplet & C$_4$F$_{10}$ & 3.0 & SNOLAB & \cite{Behnke:2016lsk} \\
PICO-60 & Bubble Chamber & C$_3$F$_8$ & 52 & SNOLAB & \cite{Amole:2019fdf} \\
SENSEI$^*$  & CCD & Si & 9.5$\times$10$^{-5}$ & FNAL & \cite{Abramoff:2019dfb} \\  
SuperCDMS$^*$ & Cryogenic & Si & 9.3$\times$10$^{-4}$ & above ground & \cite{Agnese:2018col} \\
XENON100 & TPC & Xe & 62 & LNGS & \cite{Aprile:2016swn} \\ 
XENON1T & TPC & Xe & 1995 & LNGS & \cite{Aprile:2018dbl} \\ 
XMASS & Single phase & Xe & 832 & Kamioka & \cite{XMASS:2018bid} \\ \hline \hline
\end{tabular}
\end{table}

\subsection{Annual Modulation Signature}

The only ``anomaly'' surviving since many years is the signal observed by DAMA/LIBRA. This collaboration operates a massive array of low-background NaI(Tl) crystals in the Italian Gran Sasso underground laboratory and searches for a dark matter-induced annually modulating signal. Such modulation, with period and phase consistent with the expectation from the standard halo model, has been observed in various incarnations of the project. The long-awaited results from the DAMA/LIBRA-phase~2 were published recently~\cite{Bernabei:2018yyw}. As expected from dark matter, the modulating signal is only present in single scatter events at low energies $<$6\,keV$_\tnl{ee}$. The modulation has been observed over a total of 20\,annual cycles, with a total exposure of 2.46\,t\,$\times$\,y. By now, the signal reached a statistical significance of 12.9\,$\sigma$ and could not be related to other, non-WIMP explanations.

However, there are serious concerns about the dark-matter nature of the DAMA/Libra observation: if the signal is interpreted in the standard scenarios which expect nuclear recoils signals, it is excluded by more sensitive experiments by many orders of magnitude~\cite{Aprile:2018dbl}. It is worth mentioning that the phase~2 results of DAMA, obtained with a threshold lowered from 2\,keV to 1\,keV, are also inconsistent with the results from the first phase of the same experiment when interpreted as a dark matter WIMP signal~\cite{Baum:2018ekm}. In order to avoid an interpretation bias due to a specific model, liquid xenon experiments searched for modulation signals in their data as well. These experiments are sensitive to the energy interval in which DAMA observes its signal, have a significantly lower background and were operated stably for several years. The electronic structures of xenon and iodine are very similar and the non-observation of a modulation signal thus allowed XENON100, XMASS and LUX to exclude the DAMA signal at 5.7\,$\sigma$, $\sim$2\,$\sigma$ and 9.2\,$\sigma$, respectively~\cite{Aprile:2017yea,Abe:2015eos,Akerib:2018zoq}.

Attempts to solve the long-standing discrepancy of the experimental results with so-called ``isospin-violating'' dark matter models, which favor NaI over Xe targets~\cite{Feng:2011vu}, have recently been challenged by the first results of COSINE-100~\cite{Adhikari:2018ljm}. This experiment uses a total of 106\,kg of low-background NaI(Tl) crystals which are installed inside an active veto-shield made of liquid scintillator. This allows for the reduction of the background from a 3\,keV X-ray line from $^{40}$K right in the region of interest by detecting the coincident 1.4\,MeV $\gamma$-ray in the veto. COSINE-100 did not observe an excess of signal-like events above background in a first 59.5\,d run and thus challenges the DAMA/LIBRA signal as being caused by spin-independent WIMP-nucleon interactions using the same NaI target. The first searches for an annually modulating signal with COSINE-100~\cite{Adhikari:2019off} and the similarly-sized ANAIS-112 using 112\,kg of NaI(Tl) crystals~\cite{Amare:2019jul} are both consistent with the absence of a signal, however, both need more data to draw definite conclusions on the DAMA claim.

\subsection{Nuclear Recoils from WIMP-Nucleon Scattering}

\begin{figure}[t]
\includegraphics[width=\textwidth]{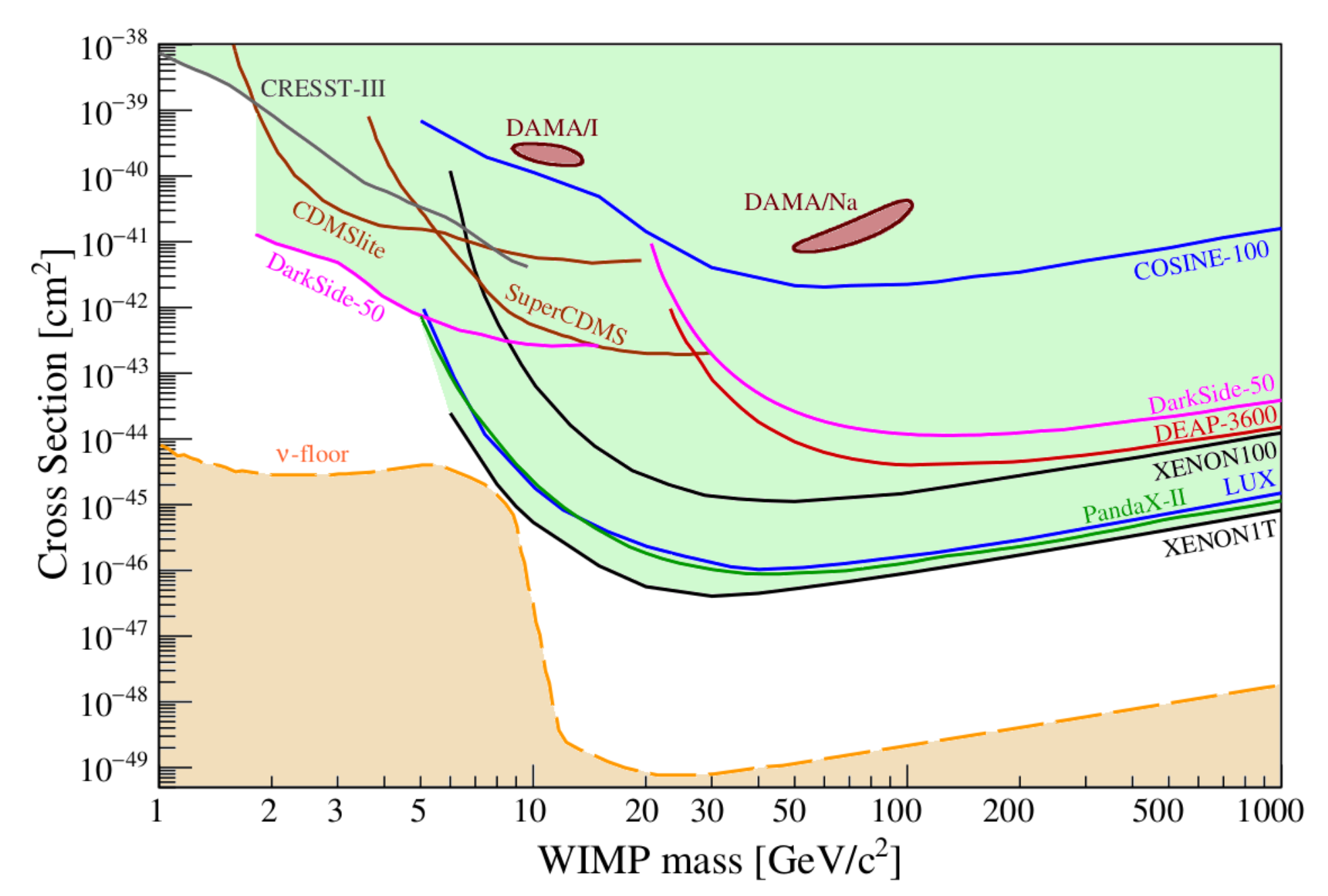}
\caption{The current experimental parameter space for spin-independent WIMP-nucleon cross sections. Not all published results are shown. The space above the lines is excluded at a 90\%~confidence level. The two contours for DAMA interpret the observed annual modulation in terms of scattering of iodine (I) and sodium (Na), respectively~\cite{Savage:2008er}. The dashed line limiting the parameter space from below represents the ``neutrino floor''~\cite{Billard:2013qya} from the irreducible background from coherent neutrino-nucleus scattering (CNNS), see Sect.~\ref{sec::floor}.} \label{fig::si}
\end{figure}

\paragraph{Spin-independent interactions} The current status of the search for spin-independent WIMP-nucleon scattering is shown in Fig.~\ref{fig::si}. (Note that only a selection of the most relevant results are shown.) Above WIMP masses of $\sim$5\,GeV/$c^2$, the strongest limits are placed by the the liquid xenon (LXe) TPCs XENON1T, LUX, and PandaX-II. The experiment with the best sensitivity to WIMPs in this mass-range is XENON1T, a LXe TPC with a 2.0\,t target. It achieved a background of only 85\,events/(t\,$\times$\,y\,$\times$\,keV$_\tnl{ee}$) in the WIMP search region (before ER rejection) and excluded spin-independent WIMP-nucleon interactions with cross sections above $4.1 \times 10^{−47}$\,cm$^2$ at 30\,GeV/$c^2$ and 90\% confidence level in a run with a 1.0\,t\,$\times$\,y exposure~\cite{Aprile:2018dbl}. The results from the liquid argon (LAr) experiments DarkSide-50 (TPC, 46\,kg target) and DEAP-3600 (single-phase, 3.6\,t target) are weaker due to their higher threshold and lower exposure. 

In a mass range from 1.8\,GeV/$c^2 \lesssim m_\chi \lesssim$\,5\,GeV/$c^2$, the most stringent exclusion limit was placed by DarkSide-50 using a LAr target depleted in $^{39}$Ar~\cite{Agnes:2018ves}. The result from a 0.019\,t\,$\times$\,y run is a based on using the ionization signal only, which allowed reducing the analysis threshold to 0.1\,keV$_{\tnl{ee}}$. The observed background of 1.5\,events/(kg\,$\times$\,d\,$\times$\,keV$_\tnl{ee}$), corresponding to $5.5 \times 10^5$\,events/(t\,$\times$\,y\,$\times$\,keV$_\tnl{ee}$), can be attributed to known background sources above $\sim$1.4\,keV$_\tnl{nr}$ (corresponding to 8\,e$^-$).

Due to their much smaller total target mass and higher backgrounds, the cryogenic experiments using Ge-crystals with ionization and phonon readout (EDELWEISS, (Super)CDMS) or scintillating CaWO$_4$-crystals with light and phonon readout (CRESST) cannot compete in the search for medium to high-mass WIMPs. However, due to their ability to reach extremely low thresholds well below 1\,keV$_\tnl{nr}$, they are very sensitive to low-mass WIMPs with masses $\lesssim$5\,GeV/$c^2$. The Germanium-based detectors SuperCDMS and EDELWEISS could improve their low-mass sensitivity by operating the detectors with a high bias voltage, converting the ionization signals into Neganov-Luke phonons. CRESST-III is currently placing the most stringent constraints below $m_\chi = 1.8$\,GeV/$c^2$~\cite{Abdelhameed:2019hmk} extending the mass range down to 0.16\,GeV/$c^2$. The result was achieved using a 24\,g~CaWO$_4$ crystal with a threshold of 31\,eV. The cryogenic crystal was operated underground and acquired an exposure of 3.64\,kg\,$\times$\,d. In this sub-GeV window competitive limits were also placed by NEWS-G, a spherical proportional counter with 60\,cm diameter and filled with a Ne$+$CH$_4$ (0.7\%) gas-mixture at 3.1\,bar (corresponding to 283\,g)~\cite{Arnaud:2017bjh}. With its low threshold of 36.5\,eV$_\tnl{ee}$ and the use of the low-$A$ gas neon the instrument was optimized to search for low-mass WIMPs.

It has recently been proposed that the reach of WIMP detectors could possibly be extended further into the sub-GeV region by exploiting the Migdal effect~\cite{Ibe:2017yqa,Dolan:2017xbu}: if the WIMP-nucleus interaction does not lead to a signal above detector threshold but the recoling atom gets excited and ionized by the process, this might lead to an additional signal which might be detected. Several results taking into account this effect were already published~\cite{Akerib:2018hck,Armengaud:2019kfj,Liu:2019kzq}, however, at the moment it is not clear how the detector response to this effect can be calibrated and whether the effect is actually present at all.  

\begin{figure}[t!]
\centering
\includegraphics*[width=0.7\textwidth]{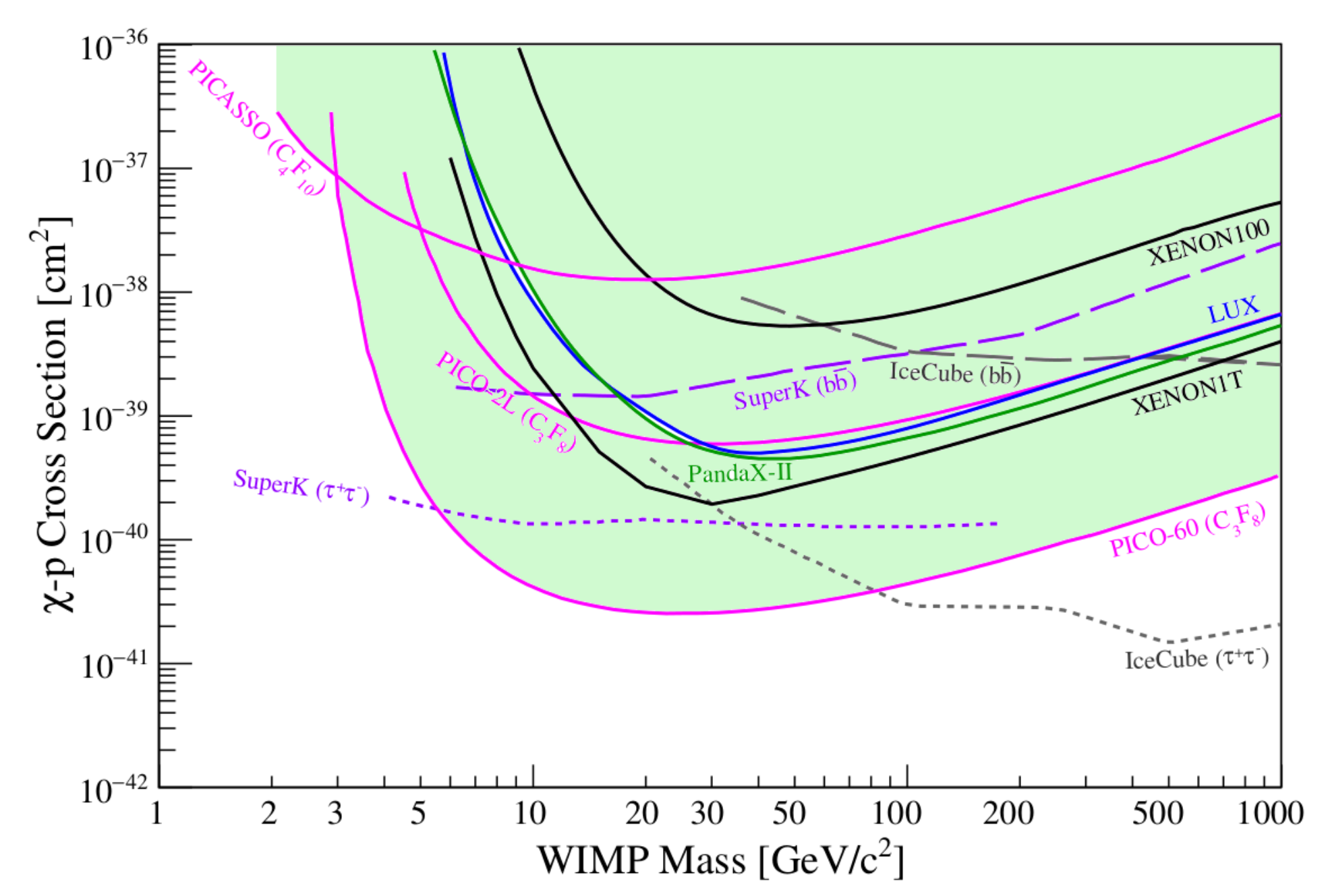}   \\ 
\includegraphics*[width=0.7\textwidth]{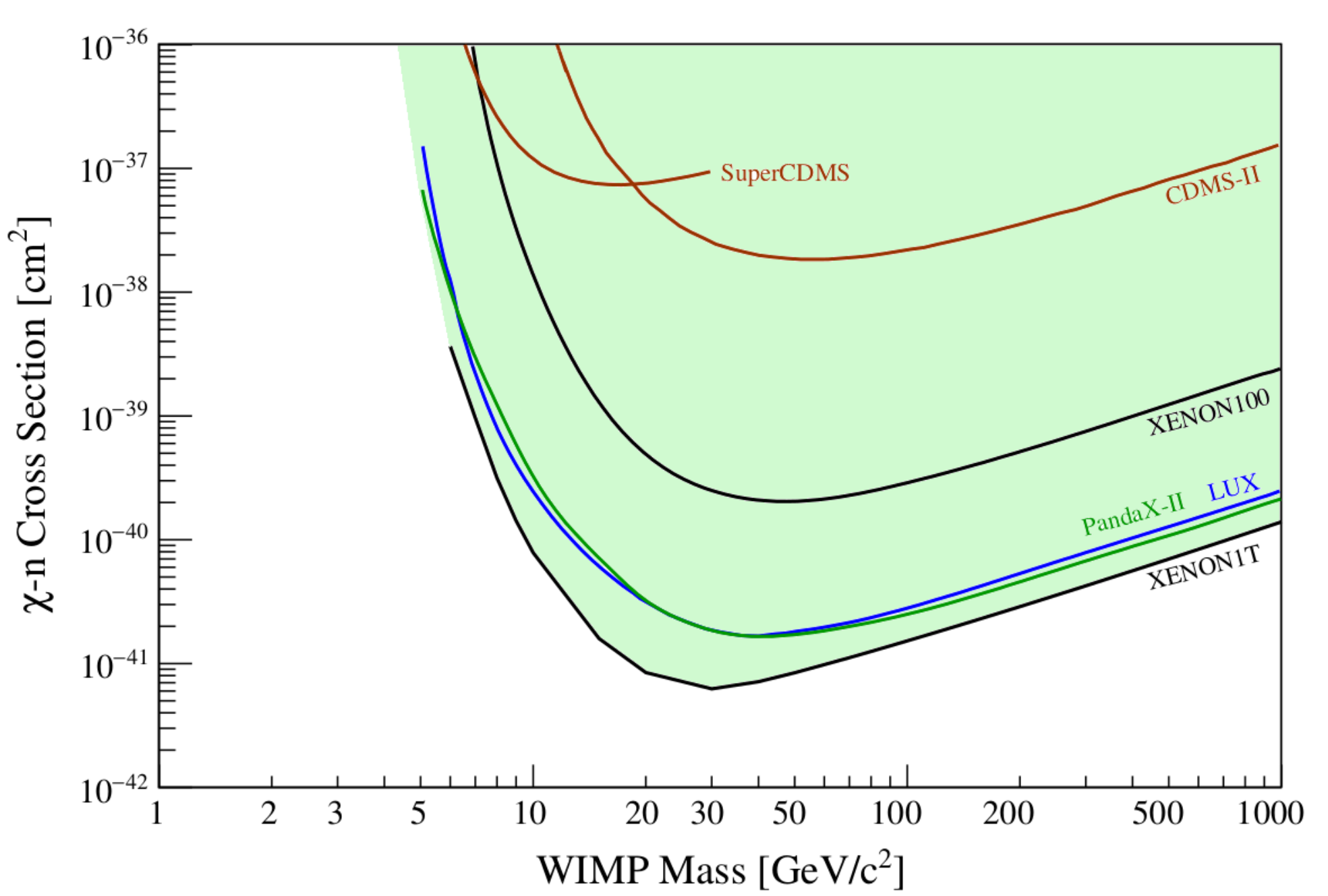} 
\caption{Current status of the searches for spin-dependent couplings. \textbf{(Top)} WIMP-proton interactions. The search is dominated by bubble chambers and superheated droplet detectors which contain the isotope $^{19}$F. The results from the much larger LXe detectors are an order of magnitude weaker. Also shown are limits from indirect searches~\cite{Aartsen:2016zhm,Choi:2015ara}. \textbf{(Bottom)} WIMP-neutron interactions. The best results are from LXe TPCs.}
\label{fig::sd}
\end{figure}

\paragraph{Spin-dependent interactions} As discussed in Sect.~\ref{sec::rates}, bubble chambers filled with targets containing the isotope $^{19}$F have the highest sensitivity to spin-dependent WIMP-proton couplings. The best limit to date is from PICO-60, operated with 52\,kg of C$_3$F$_8$ (octafluoropropane), see Fig.~\ref{fig::sd} (top). No excess of WIMP candidates was observed above the background expectation in a combined exposure of (1167$+$1404)\,kg\,$\times$\,day (with thresholds of 3.3\,keV$_\tnl{nr}$ and 2.45 keV$_\tnl{nr}$, respectively), excluding spin-dependent cross sections above $2.5 \times 10^{-41}$\,cm$^2$ for 25\,GeV/$c^2$ WIMPs~\cite{Amole:2019fdf}. At low WIMP masses between 2\,and 4\,GeV/$c^2$, the best limits on spin-dependent WIMP-proton interactions are from PICASSO which operated 32\,superheated droplet detectors with a total mass of 3.0\,kg C$_4$F$_{10}$~\cite{Behnke:2016lsk}. The experiments are probing the same parameter space as the neutrino telescopes IceCube, ANTARES and Super-Kamiokande, which constrain spin-dependent WIMP-proton scattering via dark matter capture in the Sun (and subsequent annihilation into $\tau$-leptons or $b$-quarks which decay into detectable neutrinos). At lower WIMP masses, the parameter space is also explored by ATLAS and CMS via ``mono-X'' searches.

Due to a $\sim$50\% natural abundance of xenon isotopes with unpaired neutrons, the strongest constraints on spin-dependent WIMP-neutron scattering are set again by the massive liquid xenon TPCs (LUX, PandaX, XENON) which dominate the spin-independent searches at medium to high masses, see Fig.~\ref{fig::sd} (bottom). The most sensitive result to-date is from XENON1T~\cite{Aprile:2019dbj}, re-interpreting the data already used to constrain spin-independent interactions. Argon has no stable isotopes with unpaired spins such that LAr detectors cannot contribute to spin-dependent WIMP searches. There are published results from the cryogenic Germanium experiments CDMS-II~\cite{Ahmed:2008eu} and SuperCDMS (calculated by~\cite{Marcos:2015dza} based on~\cite{Agnese:2014aze}), however, due to the significantly lower detector mass, these are not competitive.

SuperCDMS operating one Ge-crystal (600\,g) in Neganov-Luke mode (``CDMS\-Lite'') put constraints on spin-dependent WIMP-proton and WIMP-neutron interactions extending to $m_\chi=1.5$\,GeV/$c^2$~\cite{Agnese:2017jvy}, however, the results are more than six (five) orders of magnitude weaker than the best high-mass results for proton (neutron)-couplings (and thus do not fit to the axis chosen for Figs.~\ref{fig::sd}). Recent results from a 2.66\,g Li$_2$MoO$_4$ cryogenic scintillating crystal calorimeter prototype operated by the CRESST collaboration place limits down to 0.8\,GeV/$c^2$~\cite{Abdelhameed:2019szb}, however, are much weaker than the one from SuperCDMS.

\subsection{Electronic Recoils from WIMP-Electron Scattering}
\label{sec::er}

Very light WIMPs with masses in the MeV/$c^2$-range will not transfer sufficient momentum to the target nucleus to generate a detectable nuclear recoil signal. The search for such particles thus concentrates on WIMP-electron scattering. 

\begin{figure}[t]
\centering
\includegraphics[width=0.7\textwidth]{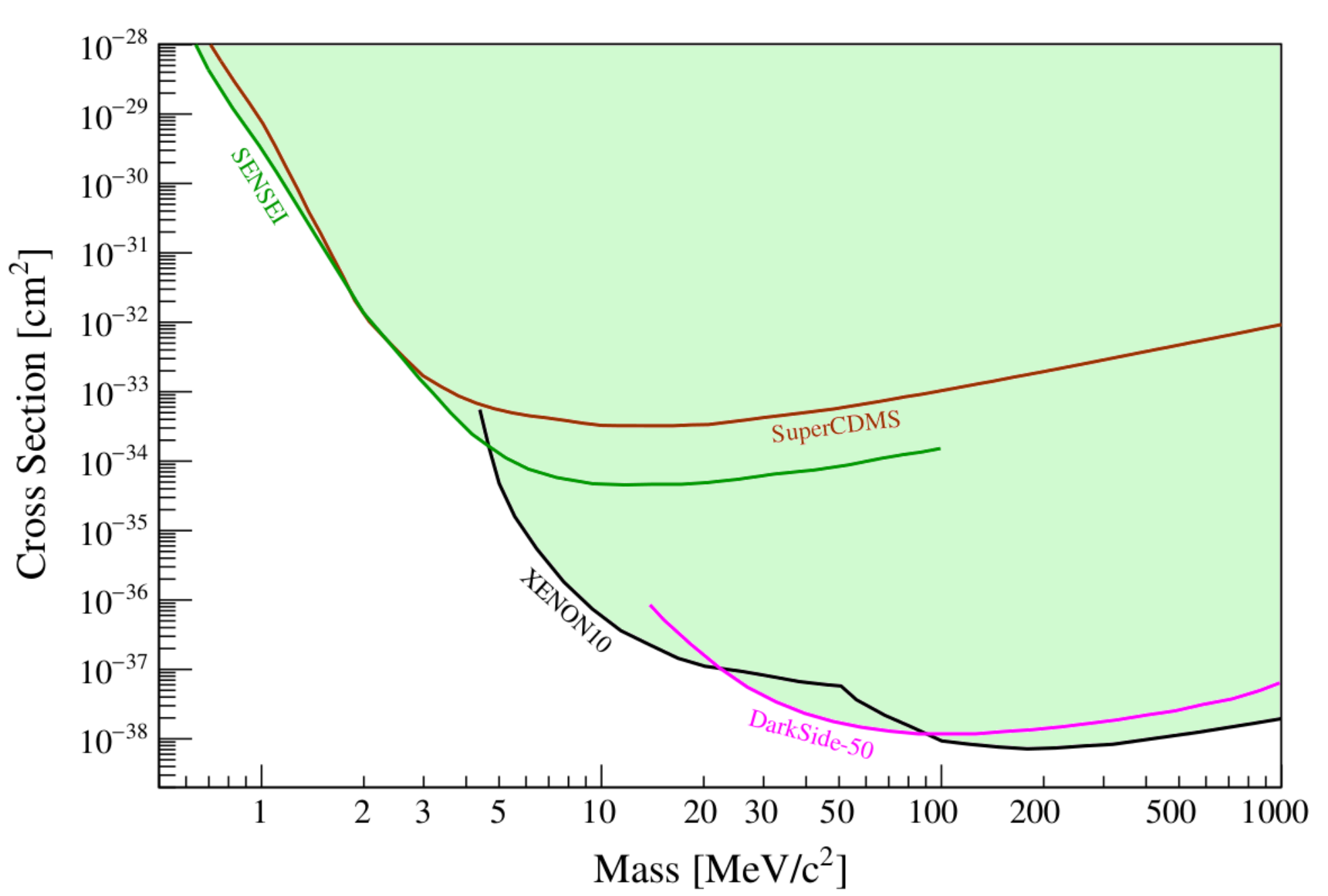}
\caption{Parameter space for MeV-scale dark matter interacting with electrons mediated by a heavy dark photon (dark matter form factor $F_\tnl{DM}=1$ independent of momentum transfer). The space is probed by experiments with sensitivities to single electrons based on solid state Si detectors (SENSEI, SuperCDMS) or dual-phase TPCs filled with LXe (XENON10) or LAr (DarkSide-50) analyzed in charge-only mode.  } \label{fig::er}
\end{figure}

Thanks to their high sensitivity to single-electron signals and the low ionization energy of 1.2\,eV, experiments using Si-based sensors are ideally suited to search for such signals. The most stringent constraints in the 0.5-5\,MeV/$c^2$ mass range come from a prototype run of the SENSEI experiment operated in the shallow MINOS cavern at Fermilab. SENSEI uses the Skipper-CCD technique featuring an ultra-low electronic readout noise of only 0.07\,electrons per CCD pixel~\cite{Tiffenberg:2017aac}. A repeated operation of the 0.0947\,g detector for 120\,s before readout led to a total exposure of 0.069\,g\,$\times$\,d and the best for WIMP-electron scattering below 5\,MeV/$c^2$~\cite{Abramoff:2019dfb}, see Fig.~\ref{fig::er}. An above-ground prototype run of a SuperCDMS detector with a resolution of 0.1\,electron-hole pairs and eV-resolution~\cite{Agnese:2018col} gave a similar result. The tiny detector masses and exposures imply that the results are expected to improve significantly once the experiments leave the prototyping stage. 
Above 5\,MeV/$c^2$, the data from dual-phase TPCs (XENON10, XENON100, DarkSide-50) analyzed in ionization-only mode provide the most stringent constraints as these detectors provide more exposure but a higher threshold than the Si detectors~\cite{Essig:2012yx,Essig:2017kqs}. 

Many more experimental efforts are currently being discussed to search for very light (eV-MeV) dark matter as an alternative to the WIMP paradigm (see, e.g.,~\cite{Battaglieri:2017aum}). These are, however, beyond the scope of this review.

\subsection{Outlook: Towards the Neutrino Floor}
\label{sec::floor}

\begin{figure}[t]
\centering
\includegraphics[width=0.7\textwidth]{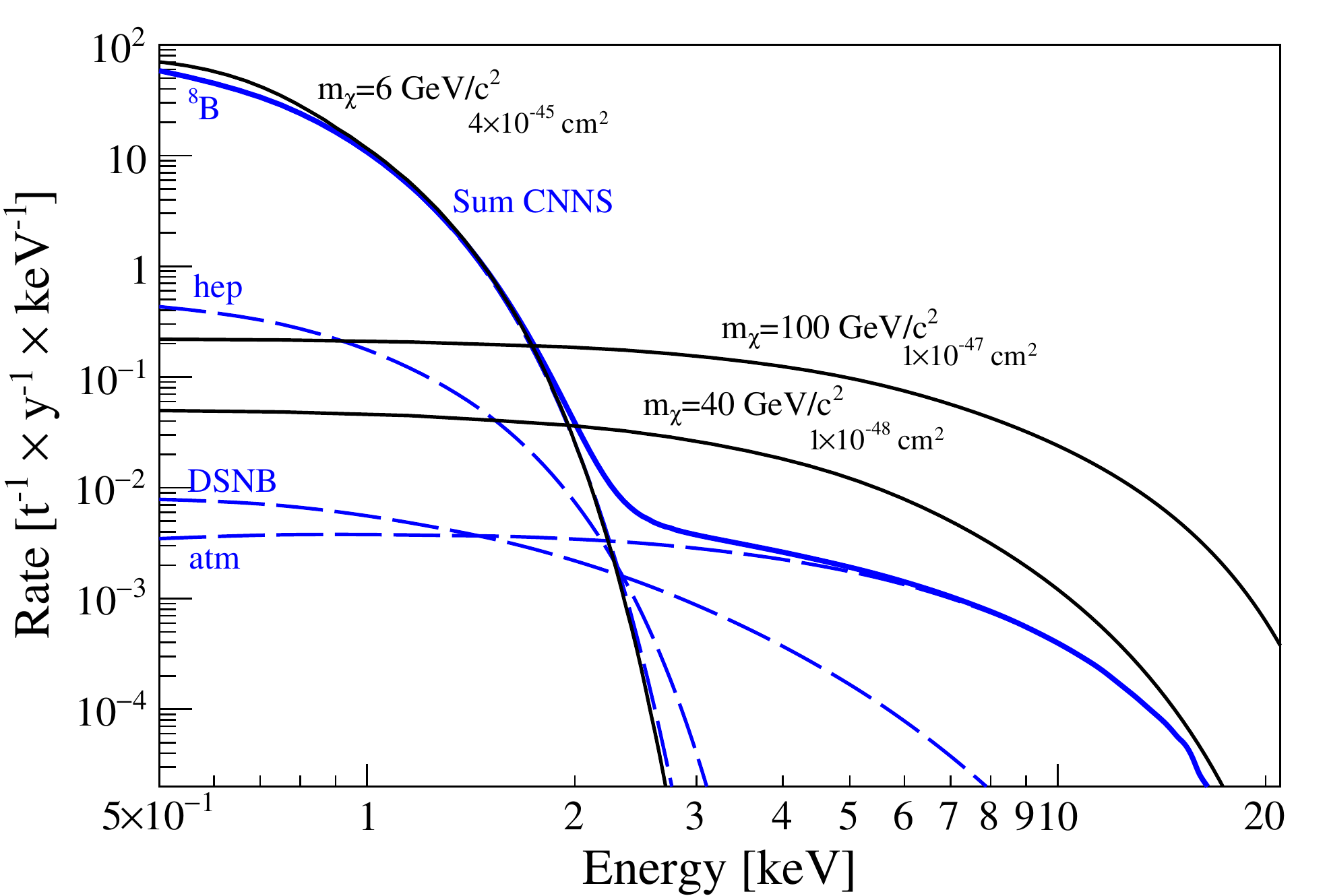}
\caption{Differential recoil spectra in xenon from WIMPs of various masses $m_\chi$ and spin-independent cross sections and from neutrinos scattering coherently off the xenon nuclei (CNNS). At low recoil energies, the CNNS sum spectrum is dominated by solar $^{8}$B neutrinos, at high energies by atmospheric neutrinos. A NR acceptance of 50\% is assumed and the detector signal is converted to the electronic recoil scale (keV$_\tnl{ee}$) using the scintillation signal only.} \label{fig::cnns}
\end{figure}

The signature for WIMP dark matter in direct detection experiments are single scatter nuclear recoils. The same signal is produced (in any detector) by coherent neutrino-nucleus scattering (CNNS) which leads to an irreducible background for the WIMP search. The cross section for neutrinos of energy~$E_\nu$ scattering coherently off a nucleus of mass~$m_N$ ($Z$~protons and $A-Z$~neutrons), resulting in a nuclear recoil of energy~$E_\tnl{nr}$, is given by~\cite{Freedman:1977xn}
\begin{equation}\label{eq::cnns}
\frac{d \sigma}{d E_\tnl{nr} } = \frac{G_F^2 m_N}{4 \pi} \left[ A-Z-(1-4 \sin^2 \theta_W) Z \right] \left(1-\frac{m_N E_\tnl{nr}}{2 E^2_\nu} \right) F^2(Q^2) \textnormal{;}
\end{equation}
$G_F$ is Fermi's constant and $\theta_W$ the Weinberg angle. As in the description of WIMP-nucleon interactions the form factor $F(Q^2)$ depends on the momentum transfer and can be described by the standard Helm form factor. The neutrinos are high-energetic $^8$B neutrinos from the Sun (dominating at low WIMP masses $m_\chi$) and atmospheric neutrinos (dominating above $m_\chi \sim 10$\,GeV/$c^2$). Neutrinos from the $hep$-process and from the diffuse supernova background (DSNB) have subdominant contributions~\cite{Strigari:2009bq}, see Fig.~\ref{fig::cnns}. This neutrino background provides the ultimate border for ``traditional'' direct WIMP searches~\cite{Billard:2013qya,OHare:2016pjy,Boehm:2018sux}, whereas directional detectors could go beyond the border provided that a significant number of events are detected~\cite{Grothaus:2014hja} (see also discussion on directional detectors in Sect.~\ref{sec::detectors}~E).

The ``neutrino floor'' plotted in Fig.~\ref{fig::si} was introduced as ``WIMP discovery limit'' in~\cite{Billard:2013qya}. It is defined by the detection of a WIMP signal at 3$\sigma$ on top of a background of 500~CNNS events above a threshold of 4\,keV$_\tnl{nr}$ in a LXe detector, assuming infinite energy resolution. Extremely large (and rather unrealistic) exposures around 5300\,t\,$\times$\,y are required to detect 500~CNNS events from atmospheric neutrinos. The cross-section where one CNNS event is expected in the dataset is about one order of magnitude higher than the neutrino floor defined in this way.

While the ton-scale LXe experiment XENON1T has already almost reached the neutrino floor at $m_\chi \sim 8$\,GeV/$c^2$~\cite{Aprile:2018dbl}, see Fig.~\ref{fig::si}, 2$-$3~orders of magnitude of cross-section are still to be explored at higher and lower WIMP masses. At low WIMP masses, where the neutrino floor is almost a factor~$10^5$ higher than at higher masses, a low detector threshold (at a moderate exposure) is the key experimental requirement. The cryogenic experiments SuperCDMS~\cite{Agnese:2016cpb} (Ge, Si-targets with standard ionization-phonon readout or with Neganov-Luke amplification) and CRESST~\cite{Angloher:2015eza} (CaWO$_4$, Al$_2$O$_3$) have presented strategies how to reach the neutrino floor in a staged approach with increasing detector mass and a significant background reduction compared to the state of the art. CRESST plans to reach the neutrino floor around 5\,GeV/$c^2$ with an exposure of 1000\,kg\,$\times$\,d of CaWO$_4$. The CYGNUS collaboration works towards 10-1000\,m$^3$-scale directional detectors to probe this region~\cite{ref::spooner2018}.

Probing the mass range $m_\chi \gtrsim 10$\,GeV/$c^2$ down to the neutrino floor requires large ``background-free'' exposures and thus massive detectors with multi-ton targets and very low background levels. Due to the exponentially falling recoil spectra, low thresholds are beneficial but not absolutely required for masses above $\sim$40\,GeV/$c^2$. This range is the realm of the massive dual-phase TPCs filled with the liquid noble gases xenon and argon. (Single-phase experiments will soon be terminated with XMASS (LXe) joining the XENON project and DEAP (LAr) joining DarkSide.) A total of four dual-phase TPCs are currently being prepared to start taking data in the next years: the LXe experiments PandaX-4T (4.0\,t active target)~\cite{Zhang:2018xdp}, XENONnT (5.9\,t)~\cite{Aprile:2015uzo} and LZ (7.0\,t)~\cite{Akerib:2018lyp} and DarkSide-20k (23\,t) operated with depleted LAr~\cite{Aalseth:2017fik}. After a few years of data taking they will be able to probe spin-independent cross sections of ${\cal O}(10^{-48})$\,cm$^2$ at the sensitivity maximum, which is still an order of magnitude above the neutrino floor. The design sensitivities of the smaller PandaX-4T and the LAr-based DarkSide-20k experiments are a bit weaker than for XENONnT and LZ, however, the achieved sensitivity of any experiment will eventually depend on the actual background level. For spin-dependent WIMP-proton couplings, the upcoming bubble chamber PICO-500~\cite{ref::VazquezTAUP} with a 500\,liter C$_3$F$_8$ target will reach similar sensitivities as XENONnT and LZ for spin-dependent WIMP-neutron scattering. 

To eventually probe heavy-WIMP cross sections at the neutrino floor, very large exposures are required. The Global Argon Dark Matter Collaboration anticipates a LAr TPC with a 300\,t fiducial mass to acquire an exposure of 3000\,t\,$\times$\,y~\cite{ref::gadmc}. Using a LXe detector, exposures of about 200\,t\,$\times$\,y and background-levels dominated by neutrinos are required~\cite{Schumann:2015cpa}. The DARWIN collaboration proposes to build a 40\,t LXe TPC with lowest backgrounds to achieve this goal. This instrument will not only be sensitive to spin-independent, spin-dependent and various inelastic WIMP-nucleon scattering interactions, but will also have a rich neutrino and astroparticle physics program~\cite{Baudis:2013qla,Aalbers:2016jon,Lang:2016zhv}.

\section*{Acknowledgments}

I would like to thank the editors for the invitation to write this review and their patience while waiting for it. This work was supported by the German Ministry for Education and Research (BMBF) through the project CRESST-XENON and the European Research Council through the ERC consolidator grant ULTIMATE.

\end{document}